\documentclass[11pt]{article}
\pdfoutput=1
\usepackage{jcappub}
\usepackage{array}
\usepackage{github_aas_macros}
\usepackage{graphicx}
\usepackage{epsfig}

\usepackage[normalem]{ulem}
\newcommand{\be}{\begin{equation}}
\newcommand{\ee}{\end{equation}}
\newcommand{\mpc}{\, {\rm Mpc}}

\newcommand{\ihmpc}{\, h\, {\rm Mpc}^{-1}}

\def\bi#1{\hbox{\boldmath{$#1$}}}
\def\bk{{\bf k}}
\def\bx{{\bf x}}
\def\br{{\bf r}}
\def\bq{{\bf q}}
\def\bs{{\bf s}}

\title{Towards optimal extraction of cosmological information from nonlinear data}  

\author[a,b,c]{Uro\v s Seljak,}
\author[a]{Grigor Aslanyan,}
\author[a]{Yu Feng,}
\author[a]{Chirag Modi}

\affiliation[a]{Berkeley Center for Cosmological Physics and Department of Physics, University of California, Berkeley, CA 94720}
\affiliation[b]{Department of Astronomy, University of California, Berkeley, CA 94720}
\affiliation[c]{Physics Department, Lawrence Berkeley National Laboratory, Cyclotron Rd, Berkeley, CA 94720}

\emailAdd{useljak@berkeley.edu, aslanyan@berkeley.edu, yfeng1@berkeley.edu, modichirag@berkeley.edu}

\abstract{One of the main unsolved problems of cosmology is how to 
maximize the extraction of information from nonlinear data. 
If the data are nonlinear the usual approach is to employ a sequence 
of statistics (N-point statistics, counting statistics of clusters, density peaks or voids etc.),
along with the corresponding covariance matrices. However, this approach is computationally prohibitive and has not been shown to be exhaustive in terms of information content. Here 
we instead develop a hierarchical Bayesian approach, expanding the likelihood
around the 
maximum posterior of linear modes, 
which we solve for using optimization methods. By integrating out the modes using perturbative expansion of the likelihood
we construct an initial power spectrum estimator, which for a 
fixed forward model
contains all the cosmological information if the initial modes are gaussian 
distributed. We develop a method to construct the window and 
covariance matrix such that the estimator is explicitly unbiased and 
nearly optimal. 
We then generalize the method to include the forward model 
parameters, 
including cosmological and nuisance parameters, and primordial 
non-gaussianity. 
We apply the method in the simplified context of nonlinear structure formation, using either simplified 2-LPT dynamics or 
N-body simulations as the nonlinear mapping between linear and nonlinear density, and 2-LPT dynamics 
in the optimization steps used to 
reconstruct the initial density modes. 
We demonstrate that the method gives an unbiased estimator of the initial power spectrum, providing among other
a near optimal reconstruction of linear baryonic acoustic oscillations. 
}

\begin{document}
\maketitle
\flushbottom

\section{Introduction}

The issue of optimal map reconstruction and optimal 
power spectrum reconstruction from a 
set of noisy and sparsely sampled data has 
received a lot of attention in the field of 
large scale structure (LSS) and cosmic microwave background (CMB)
anisotropies. Most of the work assumes the data are a linear 
transformation of the initial modes, as is the case with CMB \cite{1997MNRAS.289..285H,1997PhRvD..55.5895T,1998PhRvD..57.2117B}. 
In this case the minimum variance map solution is the well-known 
Wiener filter map \cite{1992ApJ...398..169R}. Computing Wiener filter can be expensive, 
since it requires an inverse noise weighting of the data, where in 
the cosmological context noise consists both of the actual measurement 
noise and the actual signal (the so called sampling or cosmic variance noise).
Noise covariance matrix is typically sparse (often diagonal) in the 
configuration space, and given sparsely sampled data its Fourier space representation is not diagonal. 
Signal covariance matrix is diagonal in Fourier space, and its configuration 
space representation is not diagonal. 
The sum of the two is thus not sparse in any basis, and 
computing the inverse covariance matrix using brute-force 
methods is an $N^3$ process, which 
becomes prohibitively expensive for large $N$. Alternative methods must therefore be 
used to solve for the Wiener filter map \cite{2003NewA....8..581P,ElsnerWandelt2013}. 

The reconstructed map typically has millions of data points and contains too 
much information to be useful on its own. What we want instead is the 
optimal power spectrum given the map. This is a typical hierarchical Bayesian setting, where there are many latent variables that need to be 
marginalized over, and only their priors remain. Here the latent variables are the modes and their 
prior is the power spectrum, which is a useful 
summary statistic, since for a gaussian field of initial modes 
it contains all the 
information present in the data, and the data compression is lossless. 
In the linear regime it can be computed using the optimal quadratic estimator, 
which is quadratic in the data. This requires the data to be first 
inverse covariance matrix weighted \cite{1997MNRAS.289..285H,1997PhRvD..55.5895T,1998PhRvD..57.2117B}, hence it is also 
computationally expensive. The two problems, optimal map making and optimal power spectrum, 
are connected: the optimal 
quadratic estimator can be built out of the Wiener field map reconstruction 
\cite{1998ApJ...503..492S}. 

An alternative method to optimal quadratic estimator is to use sampling methods 
to determine the power spectrum probability distribution. For example,  
a Gibbs sampling approach consists of a two step sampling procedure \cite{Wandelt:2003uk}. 
In the first one a sample map consistent with 
the data is created by adding a generalized noise realization to 
the Wiener filter map. In the second, sampling of the power spectrum is created 
consistent with the given map realization. Creating a map 
sample also requires an expensive inversion of the covariance matrix, and 
as a consequence these sampling approaches are typically slower than the optimal 
quadratic estimator, and converge particularly slowly in the high noise regime. 

These methods, while optimal in the linear regime, 
are often replaced with faster and 
less optimal methods due to their computational cost. 
Traditional power spectrum estimations of CMB and LSS 
use a linear transformation between the data and the 
model (Fourier transform, FT), and assume that the 
covariance matrix of the data 
is diagonal (which allows a fast evaluation of the 
corresponding curvature matrix of the modes). 
In LSS this is the so called FKP 
method \cite{1994ApJ...426...23F}, while in CMB it is called the pseudo-$C_l$
method \cite{1973ApJ...185..413P,2001PhRvD..64h3003W}.  
This solution is not optimal, 
because the correct weighting of the data is to multiply it with the 
inverse of the covariance matrix, an evaluation that 
scales as $N^3$ and is too costly to be performed on large data sets. 
FKP or pseudo-$C_l$
weighting is particularly poor if the noise is 
varying significantly across the survey, in which case it is optimized 
for a single value of the power spectrum (the one used in  
FKP weighting). It also fails if there is a complicated 
geometry of the survey. Moreover, the method does not 
handle optimally 
if there are modes that are contaminated and create large scale correlations
that need to be marginalized over. 
Nevertheless, these methods have become standard, 
because for linear problems one can show that under certain assumptions (typically valid 
on scales small compared to the survey size) 
the amount of information is nearly the same \cite{2004MNRAS.349..603E}. 

So far the discussion above has been about a linear mapping between the modes and the data. 
While the primary CMB is linear to a very good approximation, 
in LSS this is only valid on the 
largest scales. Most of the LSS information is contained 
on smaller, nonlinear, scales: in a 3-d LSS survey the number of modes scales as $k^3$, 
where $k\sim 1/R$ is the typical wavevector and $R$ the typical scale. 
The question of 
optimal linear map given the nonlinear data has been addressed in recent work, 
and Hamiltonian Monte Carlo (HMC) sampling approach has been used to create the map 
\cite{2013ApJ...772...63W}, as well as create samples consistent with the data for a given 
power spectrum
\cite{2010MNRAS.407...29J,2015JCAP...01..036J}. Hamiltonian sampling has an advantage over regular (e.g. Metropolis-Hastings) sampling that it can propagate far from the current sample using dynamics rather than random walk, and still the acceptance rate can be very high, both of 
which are necessary conditions for a rapid convergence in large 
dimensions due to the curse of dimensionality. However, in the current 
implementations the samples are still highly correlated, with correlation length of order 100-200 being reported 
\cite{2015JCAP...01..036J}. To create each sample one needs many Hamiltonian dynamics evaluation steps (of order 10), so the total cost of evaluation of a 
single independent realization can easily exceed 1000 calls, each being a full forward model.
It also requires knowledge of a gradient of the data model with respect to initial modes, a feature shared with the methods developed in this paper. 

In contrast to the optimal linear power spectrum analyses, there has been very little work 
in terms of extracting the optimal summary statistics given a set of noisy and 
incomplete data in the nonlinear regime. For sampling methods this may be too
difficult to solve, if HMC requires 1000 calls or more to create a single
independent realization. In the high noise regime sampling converges very 
slowly onto the correct cosmological model, because most of the mode 
power comes from the assumed power spectrum at a previous step. 
As a consequence this approach 
may not be feasible. 
Most of the focus so far has instead been to extract information 
from the two-point function of the nonlinear modes. In this case the same quadratic estimator methods discussed above can be used \cite{1998ApJ...499..555T}, but they are not optimal since there is 
information in higher order statistics as well. 
Because of this the two point function analysis is sometimes supplemented with additional statistics that 
are most likely to be complementary. Among these are the higher order correlations, 
starting with the three point function/bispectrum, various void statistics, topology statistics, 
counts of objects like clusters or other density peaks, 
and reconstruction methods, primarily focused on 
baryonic acoustic oscillations (BAO) (see e.g. \cite{2013PhR...530...87W} for a review of different probes of LSS). These approaches share the 
property that they use statistical information in addition to 
the  two point statistics
of the final field, but beyond that they differ enormously in terms of their 
motivation and scope. Combining the different statistics creates a significant 
problem of modeling their joint error distribution, since their joint 
covariance matrix cannot be computed ab-initio, but must instead be 
obtained from the simulations, which are noisy and expensive. Moreover, there is no 
guarantee that even after combining several of these statistics one will exhaust 
the information content in the data: new statistics 
are continuously being proposed (and argued to be superior by their authors), suggesting that this question is unlikely 
to be settled by this ad-hoc approach. Ideally, what is needed is an ab-initio
approach that is built with a guarantee to give a nearly optimal answer. 
This is an approach we attempt to develop in this paper. 

Since the initial density modes are assumed to be gaussian, their 
power spectrum is a lossless summary statistic, 
which should contain all the information present in the data 
(with an exception of cosmological parameters that may be required to 
map from the initial modes into the final data). The lossless nature may however break down 
in the situations of shell crossings, where there can be more than one initial solution
that maps to the same final data. 
In this paper we derive an initial linear 
power spectrum estimation, given incomplete and 
noisy data and given a nonlinear model between the initial modes and the data. 
We will show that 
as an intermediate step we will also need to derive the optimal initial density reconstruction. 
For the linear case the optimal power spectrum from the optimal density reconstruction 
has been derived in \cite{1998ApJ...503..492S}, 
and here we generalize the expressions to the nonlinear model. 
We will show that it is possible 
to cast the solution into an optimization problem, and we
develop methods to solve it efficiently. We then generalize the method to 
include forward model parameters as well. 
As a proof of principle we apply the derived 
expressions to extract the initial power spectrum from a final density 
field obtained either in an N-body 
simulation of dark matter or in a 2-LPT simulation models, and using 
a 2-LPT approximation (e.g. \cite{2002PhR...367....1B}) as
a forward model. 

\section{Statistical approach and heuristic derivation}

Following the notation of \cite{1998ApJ...503..492S}
let us assume that we measure some nonlinear observations $d(\bi{r}_i)$
at pixelized (2-d or 3-d) spatial positions $\bi{r}_i$. The data could be nonlinear 
dark matter density (although this is typically not directly observable), 
some projected dark matter density such as lensing shear or convergence, galaxy density or luminosity, Sunyaev-Zeldovich intensity etc.  
We arrange these into a vector 
$\bi{d}=\{d(\bi{r}_i)\}(i=1,...,N)$. 
Each measurement consists of a signal and a noise 
contribution, $\bi{d}=\bi{f(s,\lambda)}+\bi{d}_n$, where noise is 
assumed to be uncorrelated with the signal. 
Here $\bi{f(s,\lambda)}$ is 
a nonlinear mapping from the initial linear density modes to the final model 
prediction of the observation,
and $\bi{s}=\{s_j\}(j=1,...,M)$ 
are the underlying initial density mode coefficients that we wish to estimate.
In the statistical language $\bi{d}$ are the observed variables,
$\bi{s}$ are the latent variables 
and $\lambda$ are the forward model parameters. 
While our primary motivation for these 
is linear matter over-density $\bi{\delta}$, the formalism we develop 
is also useful for other applications such as CMB anisotropies, hence 
we keep the notation more general. 
We will 
assume these coefficients 
to be in Fourier space, where their gaussian prior can be written in a diagonal form. The modes are complex and obey $s^*(\bi{k})=s(-\bi{k})$, where 
$\bi{k}$ is the wavevector, but in our labeling of modes we will treat 
real and imaginary component as two independent modes. If the mapping 
is linear we can write $\bi{f}(\bi{s})=\bi{f'}\bi{s}$. 

We will assume the nonlinear mapping $\bi{f(s,\lambda)}$ to be computable given the initial 
density field and given some forward model parameters $\bi{\lambda}$. 
Typically this mapping will be a full N-body simulation, with some 
additional processing to produce a realistic model prediction for 
the specific observations. 
The forward model
parameters can be matter density, massive neutrinos 
or other parameters that affect the 
growth of structure, various astrophysical modeling parameters (related to 
how galaxies and baryons populate dark matter halos), and observational 
nuisance parameters (such as shear bias in weak lensing etc.). 
The mapping 
may also include smoothing (e.g. beam smoothing) or pixelization of the data. 
The true
underlying field has an infinite number of Fourier modes, but 
only a finite number of these can be estimated. 
Typically we will embed a given LSS survey into a periodic box larger than the survey, 
with zero padding in regions without the data, but in this paper we will 
simplify this to periodic box analysis.

The modes $\bi{s}$ are latent variables with a prior of their own. This prior is 
parametrized as a multivariate gaussian, 
with covariance matrix
$\bi{S}=\langle \bi{ss}^{\dag}\rangle$, which is assumed to be diagonal 
in Fourier space. We will assume the power spectrum depends on parameters 
$\bi{\Theta}$, which will typically be bandpowers, but this will later be generalized
to any parameters that change the power spectrum. These bandpowers can also have a 
prior (a hyperprior in the language of hierarchical Bayesian models), but in this 
paper we will assume their prior is flat, and we will not even impose positivity, 
since these are summary statistics expected to be used in a later analysis of cosmological 
parameters. 

The noise vector
$\bi{d}_n=\{d_{n,i}\}(i=1,...,N)$ 
is parametrized with the 
noise covariance matrix $\bi{N}=\langle \bi{d}_n\bi{d}_n^{\dag}\rangle$.The i-th diagonal element $N_{ii}$ corresponds to the noise variance $N({\bf r}_i$ at the spatial position ${\bf r}_i$. 
This noise matrix is assumed to be known, uncorrelated with the signal
and diagonal (or sparse) in real space, so that any operations involving noise matrix scale as $O(N)$ rather than $O(N^2)$ or steeper.  
For simplicity we will also assume noise is gaussian distributed, but this 
can be generalized to other probability distributions. It is 
reasonably appropriate for weak lensing, where by central limit theorem 
averaging over shape noise produces approximately gaussian noise. 
Outside the survey mask (which can include holes inside the survey) we will assume $N(\bi{r}_i)=\infty$
and assign $d(\bi{r}_i)=0$ (although assigning 
any other value would be just as good). 

The goal of parameter inference is to derive the posterior distribution of 
parameters $\bi{\Theta}$ and $\bi{\lambda}$ given their prior and 
the data $\bi{d}$. 
In some cases, such as linear model, one can write analytic expression 
for the likelihood $L(\bi{d} | \bi{\Theta},\bi{\lambda})$, but its 
explicit evaluation requires inversion and trace or 
determinant of a very large non-sparse matrix, both of which are $O(N^3)$ operations, 
which becomes too expensive when the size of the data becomes large. 
In the nonlinear case even writing down the posterior is a major 
challenge. We will argue that it is easier to solve the problem in a typical 
hierarchical Bayesian model approach, 
by working with the latent variables $\bi{s}$ and solving for these first, 
then marginalizing over them to obtain the likelihood and the posterior of parameters (which are the same if the priors are flat, as we will assume here). This 
approach perhaps seems counter-intuitive, since the dimension of 
the latent variables $M$ is comparable to the data $N$. However, the 
major simplification in the linear case is that there are no large 
matrix inversions required for the solution to be obtained. For the 
nonlinear case we will argue that finding a good solution at or close to the 
global minimum is achievable even in a large number of dimensions, and 
the problem cannot even be solved in the absence of latent variables. By writing 
a full probabilistic model in terms of conditional probabilities between 
individual variables, in this case latent variables $\bi{s}$ conditional on 
$\bi{\Theta}$, 
and data $\bi{d}$ conditional on $\bi{\lambda}$ and $\bi{s}$, 
we are able to solve for the posterior of $\bi{\Theta}$ and $\bi{\lambda}$
given the data $\bi{d}$ (combining with the (hyper)priors on $\bi{\Theta}$ and $\bi{\lambda}$). 

One can write the joint probability of signal $\bi{s}$
and noise $\bi{d}_n$ as a product of individual probabilities, under
the assumption that they are uncorrelated,
\begin{equation}
P(\bi{s},\bi{n})=P_s(\bi{s})P_n(\bi{d}_n)=(2\pi)^{-(M+N)/2} \det(\bi{SN})^{-1/2}
\exp\left(-{1 \over 2}
\left[\bi{s}^{\dag} \bi{S}^{-1}\bi{s}+\bi{d}_n^{\dag} \bi{N}^{-1}\bi{d}_n\right]\right).
\label{psn}
\end{equation}

The Bayes theorem can be applied to $\bi{s}$, $\bi{\Theta}$ and 
$\bi{\lambda}$ to 
obtain their posterior
\begin{equation}
P(\bi{s,\Theta,\lambda}|\bi{d}) 
\propto P_n(\bi{d} -\bi{f}[\bi{s,\lambda}]) P_s(\bi{s}| \bi{\Theta})
P(\bi{\Theta,\lambda}).
\label{psn2}
\end{equation}
The first term $P_n$ is the probability distribution of the noise for 
some set of observations $\bi{d}$, $\bi{d}_n=
\bi{d}-\bi{f}(\bi{s,\lambda})$, while 
the second $P_s$ is the prior for $\bi{s}$ given $\bi{\Theta}$ (or, 
equivalently, $\bi{S}$). 
The last term $P(\bi{\Theta,\lambda})$ is the prior on the initial power spectrum parameters 
$\bi{\Theta}$ and on forward model parameters 
$\bi{\lambda}$. We will assume this prior is flat, 
so that we have no prior information on them. This is because we want the 
result of the analysis to summarize the information from the given data, without inclusion of external data. This is easy to modify at a later 
stage if 
combining different data sets. Similarly, we will not be concerned with the normalization of the posterior, so we simply use the proportionality symbol in equation \ref{psn2}. We will fix 
the normalization at the end using gaussian approximation for the posterior. 

Maximizing the posterior in terms of $\bi{s}$
at a fixed $\bi{\Theta}$ and $\bi{\lambda}$ 
gives $\hat{\bi{s}}$, the so called maximal a posterior (MAP),
which is also the Wiener filter map in the linear case. 
We will use optimization methods with analytic gradient evaluation 
to perform this step.
Gradient based optimization methods
can be very efficient at finding the maximum, but typically do not 
provide a reliable curvature matrix (Hessian or its inverse).
We will develop a method where the curvature matrix
of $\bi{s}$ around $\hat{\bi{s}}$ 
is not needed in the construction of the final 
solution, so that any optimization method can be used. 

Ultimately the optimal map is not what we really care about. What 
we want is the 
posterior of model parameters
$\bi{\Theta}$ and $\bi{\lambda}$ independent of latent variables $\bi{s}$. 
One way to do this would be to use sampling of both $\bi{s}$, $\bi{\Theta}$ and $\bi{\lambda}$, but with 
$M\gg 10^6$ dimensions this can be
extremely expensive. 
Another approach would be to
maximize the posterior with respect to all the parameters at once. This
approach is strictly not valid when the latent variables depend on the parameters we 
wish to determine: maximization is not the same as marginalization. In addition, 
this approach has technical difficulties related to the fact that 
for the parameters we care about we also need a very accurate curvature 
matrix and its log determinant, which is not provided by the optimizer. 
In this paper we instead perform an analytic marginalization over the latent 
variables. 

In our approach, we work with large dimensions of the 
modes, $M \gg 10^{6}$.  
We will assume that the same maximum is always found. 
To be more precise, we will argue that many posterior maxima can exist and do not change
the nature of the solution as long as they are all similar to each other in a 
well defined sense discussed in section \ref{sec3}). 
This assumption can be explicitly proven in the linear 
regime where the loss function is convex \cite{1998ApJ...503..492S}. 
In the nonlinear case it depends on 
the nature of the problem,
but cannot be proven generally. For example, 
in the strongly nonlinear regime, 
inside virialized
regions, satellites on orbits can yield several identical phase space 
configurations. While
this degeneracy will be broken
by the gaussian prior on the modes, it still suggests local maxima in posterior (even if not of equal height). We will proceed not ignoring this issue, but assuming it can be calibrated 
out under the assumption that the maxima are always quantitatively similar. 

We will thus analytically integrate over $\bi{s}$ 
\begin{equation}
P(\bi{\Theta,\lambda}| \bi{d})
=\int d^M\bi{s} P(\bi{s,\Theta,\lambda}|\bi{d}),
\end{equation}
performing this marginalization around the MAP solution $\hat{\bi{s}}$. 
We will perform a perturbative expansion of the posterior 
around MAP, which we can integrate over. This produces 
two terms, one is simply the value of the loss function in the 
exponent of equation 
\ref{psn2} at the MAP position, 
and the second is the volume of posterior, roughly given by the 
determinant of the curvature matrix at that 
position (and further increased by the higher order moments). 
However, we will never explicitly perform this 
integral, and instead develop a method to determine the resulting 
volume element using simulations, so in this sense our approach 
does not rely on perturbative expansion. 

Once we have $P(\bi{\Theta,\lambda}| \bi{d})$ we can perform 
its maximization with respect to $\bi{\Theta,\lambda}$
to find their peak posterior solution, i.e. the
maximum likelihood (since we assume flat prior). By expanding the log posterior to 
second order we also obtain the curvature matrix under the Laplace approximation. 
We will use Newton's
method to find the maximum likelihood solution and then construct an 
explicitly unbiased estimator out of it. In this paper we focus primarily on 
bandpowers for $\bi{\Theta}$, and we argue that  
quadratic expansion of their log likelihood is likely to be sufficient, 
but we also develop the method for other parameters, such as
forward model parameters. 

In the past work on linear problem \cite{2003NewA....8..581P}, the evaluation 
of the curvature matrix has proven to be 
the most expensive part of the problem. In this paper we develop a novel 
method to evaluate an approximation to its ensemble average, the 
Fisher matrix. We use data simulation(s)  
and investigate their response to 
the change in parameters, taking advantage of the fact that in Newton's 
method the response matrix is also the curvature matrix. This enables us 
to evaluate this term much faster than otherwise possible. In the 
linear case this approach remains Bayesian, since the curvature matrix 
equals the Fisher matrix. In the nonlinear case this is no longer 
the case,
but we will argue that the difference are likely to be small due to the 
large number of modes. 

It is worth emphasizing that if the initial equations \ref{psn} and 
\ref{psn2} are exact in 
a probabilistic sense, and if all the steps we described above are 
performed exactly, then the final solution is 
optimal, i.e. we obtain true posterior distribution of the 
parameters given the data. In this paper we will discuss in detail 
which approximations we are invoking to solve these equations and what their 
impact may be on the optimality of the final result. 

\subsection{Heuristic derivation}

The simplest power spectrum estimate one can make out of MAP is to square the modes $\bi{\hat{s}}$ and average them within the bandpower bin. 
This does not lead to an unbiased estimator, but intuitively this must be the path to the correct procedure, since it is built from the minimum variance estimator of the initial modes, which in some sense is the best we can do. 
This procedure has indeed been formally proven in the linear case \cite{1998ApJ...503..492S}. 
In the next sections we
present a nonlinear version of this statement. Before proceeding we 
give a heuristic derivation of our procedure following the arguments in
\cite{1998ApJ...503..492S}.  

In general an optimal two point function (e.g. power spectrum) analysis for the linear case requires first inverse
generalized noise weighting of the data, after which one adds up all 
pair products of these inverse noise processed 
data, weighted by the expected response of each pair product to the 
specific power spectrum parameter (e.g. bandpower). 
Inverse weighting by the generalized noise means weighting by the inverse of the 
full covariance matrix, which consists of noise variance and signal (sampling) variance, and is in general 
not diagonal in any basis. In the nonlinear case the generalized 
noise matrix cannot even be defined a priori since it depends on the 
solution. However, 
finding MAP of initial modes $\hat{\bi{s}}$ 
achieves this inverse generalized noise weighting. This may still 
not be sufficient for an optimal analysis, but going beyond it 
would require doing a full likelihood analysis in $M$ dimensional space, 
which we would like to avoid if possible. 
So our strategy will be 
to use this solution to construct the linear power spectrum. 

While minimum variance mode reconstruction $\hat{\bi{s}}$ 
has been inverse generalized noise weighted, its square cannot 
give an unbiased estimator of the power spectrum, and needs to be 
corrected. First of all, $\hat{\bi{s}}$ contains noise contribution, 
and upon squaring it this will give a term that needs to be 
removed. Second, minimum variance modes $\hat{\bi{s}}$ only agree 
with truth in the absence of noise and for a complete coverage, but 
are otherwise reduced in amplitude and mixed with each other. So one 
needs a mixing matrix to correct for these effects. 

Heuristically, 
we can write an estimator for power $S(k)$ at the single mode $k$ as
\begin{equation}
(\bi{F}\hat{\bi{S}})_{k}=
\frac{\vert \hat{s}(k) \vert ^2}{2S_{\rm fid}(k)^2} -b_{k},
\label{wfmv0}
\end{equation}
where $\bi{F}$ is the mixing matrix between the modes 
and $b_k$ is the noise bias term. This term can be determined by 
requiring that the equation above is unbiased at some fiducial 
power spectrum $S_{\rm fid}$. 
We have divided by the sampling variance of the power spectrum 
for a single mode $2S_{\rm fid}(k)^2$ (to be evaluated at some fiducial power 
spectrum). In the limit where there is no noise the latter is the variance 
of that mode power (note that we are imagining one independent 
mode per each $\bi{k}$, while 
in practice these are complex modes with two independent mode realizations, 
but where the mode at $\bi{k}$ is a complex conjugate of the mode at 
$-\bi{k}$). So in this limit this gives the 
inverse variance weighting of the power spectrum, where the variance 
is given by the signal (sampling variance). 
Noise variance is accounted for by the reduced value of $\hat{s}_k$ relative 
to no noise case, an outcome of the minimum variance reconstruction of MAP
$\hat{\bi{s}}$. As shown below, 
the above expression correctly weights by the inverse 
generalized noise of the power. 

We will interpret
$\bi{F}$ as the matrix that determines the mixing between 
the different power spectrum estimators (where in the expression above each 
mode leads to an independent power spectrum estimator). We will 
determine both $\bi{F}$ and $\bi{b}$ using simulations around 
some fiducial model, enforcing that
equation \ref{wfmv0} gives an on average unbiased estimation for any small 
power spectrum variations around some 
fiducial model. 
We will 
show that we can
interpret $\bi{F}$ as the 
Fisher matrix, i.e. the ensemble averaged inverse 
covariance matrix for the estimators, 
as a consequence of Newton's method. 
Since equation \ref{wfmv0} leads to an estimator for each mode it is customary 
to average them into bandpowers, by adding up estimators within a 
certain range of wavevectors. 

The plan of the next sections 
is as follows. In section \ref{sec2} we construct the 
minimum variance solution for $\bi{s}$. We will assume 
$\bi{\Theta}$ and $\bi{\lambda}$ 
are fixed and we will not carry their dependence. 
In section \ref{sec3} we marginalize over $\bi{s}$ and construct 
maximum likelihood solution for $\bi{\Theta}$, assuming $\bi{\lambda}$ 
is fixed. In section \ref{sec4} we also allow variation of 
$\bi{\lambda}$, constructing joint maximum likelihood solution of 
both $\bi{\Theta}$ and $\bi{\lambda}$, and providing a general solution. 
We then extend this to optimal higher order estimators such as 
bispectrum. In section \ref{sec5} we apply 
the method to a simple but nontrivial nonlinear example, which is followed by 
conclusions in section \ref{sec6}. 

\section{Maximum a posteriori (MAP) initial mode estimator}
\label{sec2}

In this section we work out the minimum variance modes $\bi{s}$ at fixed $\bi{\Theta,\lambda}$ around some fiducial model, so we will not 
include their dependence in the expressions below. For simplicity we will 
use $\bi{S}$ for the fiducial model. 
The conditional posterior
probability for $\bi{s}$ given the data and prior $P_s(\bi{s})$ is $P(\bi{s} \vert \bi{d})
\propto P_s(\bi{s})P_n(\bi{d-f(s)})$, which from equation \ref{psn} is
\be
P(\bi{s} \vert \bi{d})= (2\pi)^{-(M+N)/2} \det(\bi{SN})^{-1/2}\exp\left(-{1 \over 2}
\left\{\bi{s}^{\dag} \bi{S}^{-1}\bi{s}+\left[\bi{d-f(s)}\right]^{\dag} \bi{N}^{-1}\left[\bi{d-f(s)}\right]\right\}\right).
\label{lik}
\ee

The MAP solution of the initial modes can be obtained by maximizing
the posterior of $\bi{s}$ at a fixed $\bi{S}$ and $\bi{N}$, 
which is equivalent to minimizing the loss function $\chi^2$, 
\be
\chi^2(\bi{s})=\bi{s}^{\dag} \bi{S}^{-1}\bi{s}+[\bi{d-f(s)}]^{\dag} \bi{N}^{-1}[\bi{d-f(s)]},
\label{chi2_eq}
\ee
with respect to all coefficients $\bi{s}$. 

The problem of finding the minimum variance field $\bi{s}$ can thus be recast into solving 
a minimization problem of a nonlinear function $\chi^2$, which is  an optimization problem.
A generic class of optimization models is based on second order Newton's method, which expands the nonlinear function to a 
quadratic order around a point $\bi{s}_m$, 
\be
\chi^2(\bi{s})=\chi^2_0+2\bi{g}^{\dag}\Delta \bi{s}+\Delta \bi{s}^{\dag}\bi{D}\Delta \bi{s},
\label{chi2}
\ee
where $\Delta \bi{s}=\bi{s}-\bi{s}_m$. 
We can introduce the gradient matrix around $\bi{s_m}$ as
\be
f'_{ij}={\partial f(\bi{s_m})_i \over \partial s_j}.
\label{rij}
\ee
Response function depends on the position $\bi{s_m}$, but we will not keep 
that index explicitly in our expressions. For linear models the matrix $\bi{f'}$
is a constant independent of position. 
The gradient of the cost function is
\be
\bi{g}={1 \over 2} {\partial \chi^2 \over \partial \bi{s}}=\bi{S}^{-1}\bi{s_m}-\bi{f'^{\dag}(s_m){N}^{-1}[d-f(s_m)}].
\label{gder}
\ee
We want to find a solution where $\bi{g}=0$, and we will denote the solution as $\hat{\bi{s}}$, 
\be
\bi{g}=\bi{S}^{-1}\hat{\bi{s}}-\bi{f'^{\dag}(\hat{s}){N}}^{-1}[\bi{d}-\bi{f}(\hat{\bi{s}})]=0.
\label{gder1}
\ee
Note that $\bi{f'}$ is the derivative 
of the forward model at 
every position with respect to every initial mode. This can be very expensive to 
compute for complicated nonlinear models since it requires back-propagation, 
and is typically not part of the standard N-body simulation codes. 

The curvature matrix $\bi{D}$ equals one half of Hessian matrix and can be written as 
\be
\bi{D}={1 \over 2} {\partial \chi^2 \over \partial \bi{s} \partial \bi{s}}=
\bi{S}^{-1}+\bi{f'}^{\dag} \bi{N}^{-1}\bi{f'}+\bi{f''}\bi{N}^{-1}[\bi{d}-\bi{f}(\bi{s_m})].
\label{D1}
\ee
The last term above contains a second derivative of the $\chi^2$, 
$\bi{f''}=\partial^2{f}/\partial \bi{s}\partial \bi{s}$,  and is often neglected when performing 
optimization, 
because it fluctuates around zero uncorrelated with the model and hence 
tends to average to zero. It is also often small (or exactly zero for linear models), and its inclusion 
may make the optimization unstable \cite{1992nrca.book.....P}. 
We will also not keep it in our expressions below, and the procedure becomes 
the so called Gauss-Newton method. Keeping or dropping this term has no impact on the final solution of the optimization. 

Minimizing $\chi^2$ equals setting its gradient with respect to $\Delta \bi{s}$ 
to zero, 
\begin{equation}
\frac{\partial \chi^2(\bi{s})}{\partial \Delta \bi{s}}=0,
\end{equation}
which from equation \ref{chi2} gives 
\be
\Delta \bi{s}=-\bi{D^{-1}g}. 
\label{shat}
\ee
If we had a perfect inverse curvature matrix and the system was linear this would find the solution 
where $\bi{g}=0$.
In nonlinear Newton's method one takes this solution as a starting point, performing a 
line search in the direction of $\bi{\Delta s}$ to find a suitable new position 
$\bi{s}_{m+1}$, which reduces the cost function 
$\chi^2$ along the line,
\be
\bi{s}_{m+1}-\bi{s}_{m}=-\alpha \bi{D^{-1}g}. 
\label{shat1}
\ee
We start with $\alpha=1$. 
If this reduces the cost function we accept it as the next iteration, 
otherwise we use bisection method to determine a new $\alpha$, repeating 
the procedure (if no solution is found we switch to regular steepest 
descent method). 
This
line search does not require a new evaluation of the derivatives, but it does require a full forward model
evaluation of $\bi{f(s_m+\Delta s)}$ for each $\alpha$, and hence can be expensive. 
The procedure is repeated until one converges to the minimum $\bi{\hat{s}}$. 

As stated above in linear models one can perform this step only once, starting from $\bi{s}_0=0$, 
to give
\be
\bi{\Delta s}=\hat{\bi{s}}=\bi{D^{-1} f'^{\dag}{N}^{-1}d}.
\label{lwf}
\ee
This is the Wiener filter solution, which minimizes the variance \cite{1992ApJ...398..169R}. Note that even though the solution is analytic, 
it requires inversion of a large matrix $\bi{D}$, which is an $M^3$ process
and becomes prohibitively expensive for large $M$. Optimization methods 
discussed here can be the best choice even for linear models. 

At the position of the minimum one can write the conditional probability of $\bi{s}$ given the data 
as
\begin{equation}
P(\bi{s} \vert \bi{d}) \propto \exp\left(-{1 \over 2}[\bi{s}-\hat{\bi{s}}]^{\dag}
\bi{D}[\bi{s}-\hat{\bi{s}}]\right),
\label{psd}
\end{equation}
where $\hat{\bi{s}}$ is the value of $\bi{s}$ found at the minimum. The covariance
matrix of residuals is given by the inverse of curvature matrix 
\begin{equation}
\langle [\bi{s}-\hat{\bi{s}}]^{\dag}
[\bi{s}-\hat{\bi{s}}]\rangle=
\bi{D}^{-1} = \left(\bi{S}^{-1}+\bi{f'}^{\dag} \bi{N}^{-1}\bi{f'}\right)^{-1},
\label{D}
\end{equation}
where the gradient 
$\bi{f'}$ is computed at the MAP $\hat{\bi{s}}$ and we dropped 
$\bi{f''}\bi{N}^{-1}(\bi{d}-\bi{f})$ term, which fluctuates around zero.  

Function optimization (minimization/maximization) is a 
large field of research, and many different methods 
have been developed for a wide range of applications \cite{Nocedal2006NO}. 
Here we will work in a large number of dimensions (with both $M$ and $N$
expected to be in millions or more), and a brute force Hessian 
inversion in equation \ref{shat} is not feasible. 
Two of the popular classes of optimization solvers
are conjugate gradient methods (often coupled to a 
preconditioner), which do not require 
a construction of curvature matrix (or its inverse) at all, and 
quasi-Newton's methods, which approximate the inverse Hessian, of which limited 
memory BFGS (L-BFGS) is the most popular 
implementation \cite{1992nrca.book.....P}. 
We have used Gauss-Newton's approach in the derivation above to highlight the 
analogy with linear algebra derivation of \cite{1998ApJ...503..492S}, 
but ultimately the choice 
of optimization method will be determined by its effectiveness. In our tests we 
have found L-BFGS to be faster than conjugate gradient, but this could depend on the application and implementation. 
The derivation above does not include normalization, 
but in practical applications we have found to be 
useful to rescale (normalize) and work with the modes $\bi{u}=\bi{s/S}^{1/2}$.

\section{Minimum variance initial power spectrum estimator}
\label{sec3}

While knowing the MAP solution $\hat{\bi{s}}$
is useful to create a map of the universe, this is still too much information 
for most applications. What we want to know is the statistical distribution of 
the linear modes, and since these are assumed to be gaussian a suitable 
summary statistic is their power spectrum. Basically, we want to know the 
parameters $\bi{\Theta}$, which are priors for $\bi{s}$ and marginalize over $\bi{s}$. In this section we address the dependence on parameters $\bi{\Theta}$, 
while still keeping parameters $\bi{\lambda}$ fixed at their fiducial 
values (and we will not carry their dependence). 

Formally we want to maximize the posterior probability  of the data as a function of some power spectrum parameters $\Theta_l$ that determine the 
initial mode power spectrum $\bi{S}$, 
independent of the underlying field $\hat{\bi{s}}$. 
We can introduce a 
derivative matrix $\bi{\Pi}_l$ around some fiducial power spectrum $\bi{S}^{\rm fid}$, defined as
\begin{equation}
\left[{\partial \bi{S} \over \partial \Theta_l}\right]_{\bi{S}^{\rm fid}}=\bi{\Pi}_l,
\label{pi}
\end{equation}
which is to be evaluated at the fiducial model. 
In terms of this we can write
\be
\bi{S}=\bi{S}^{\rm fid}+\sum_l\Delta \Theta_l\bi{\Pi}_l.
\ee
For linear dependence of $\bi{S}$ on $\bi{\Theta}$ and bandpowers
we can use 
\be
\bi{\Pi}_l={\bi{S}_{\rm fid} \over \Theta_l}, 
\label{pilin}
\ee
i.e. $\langle s_{k_l}s_{k_l}^* \rangle=\Theta_l\Pi_l(k_l)$. In this case the derivative matrix takes us from $\Theta_l$, which is the power spectrum value representative over a bin, to $\bi{S}$ that is the power spectrum. 
For concreteness we will initially bin the power spectrum into bins, 
summing over all the $K_l$
modes $\{s_{k_l}\}(k_l=1,...,K_l)$. These are the modes that
contribute to $l$ bandpower of the power
spectrum parametrized with $\Theta_l$. 
For a narrow bin the derivative matrix $\bi{\Pi}_l$
consists of ones along the diagonal 
corresponding to the $K_l$ modes and zeros otherwise.
For broader bandpower bins 
the derivative matrix can allow for any expected variation of 
the power within the bin. For example, if we only have a single bin 
which covers the entire range of modes then the corresponding $\Theta_1$
is a power spectrum amplitude chosen at some wavemode $k_f$ (or some integral 
over the modes, such as $\sigma_8^2$ normalization), 
while the derivative matrix has the shape of a fiducial power spectrum, i.e. 
$\bi{\Pi}_1=\bi{S}_{\rm fid}/\Theta_1$. 

\subsection{Analytic marginalization with perturbative approach}

In the case of power spectrum estimation
one wants to marginalize the full posterior 
over the parameters $\hat{\bi{s}}$. Since a brute force 
approach with sampling is too expensive, we choose to approximate it 
as a multi-variate gaussian as in equation \ref{psd}. An implicit 
assumption in the procedure is that there is one global minimum, to 
which optimization converges regardless of the starting point. 
This is a major assumption and may not be valid, either because the global minimum is not reachable (because optimization is stuck in a 
local minimum), because the correct solution does not correspond
to the global minimum, or because the posterior around the global 
minimum is not well described as a multi-variate gaussian. 
Here we will assume this approach is justified but there are no 
guarantees and indeed on very small scales the methodology may 
need to be modified. 

Under this assumption our strategy is to marginalize over the modes, by 
writing it as a multi-variate gaussian
around the maximum, and then complete the square.  
Note that the MAP $\hat{\bi{s}}$ and its 
covariance $\bi{D}$ 
is implicitly a function of the data, the forward model parameters and the 
power spectrum bandpower parameters $\bi{\Theta}$ we wish to estimate. In \cite{2003PhRvD..68h3002H} the resulting 
expression was called a grand likelihood $L(\bi{d} | \bi{\Theta})$. The full problem can be 
solved if we linearize the response around the MAP initial mode estimator $\hat{\bi{s}}$ and perform analytic 
integration under gaussian approximation. We begin with equation \ref{lik}, expanding around $\bi{s}=\hat{\bi{s}}+\delta \bi{s}$,
\begin{eqnarray}
&&P(\bi{s},\bi{d} \vert \bi{\Theta})=(2\pi)^{-(M+N)/2} \det(\bi{SN})^{-1/2} \times \nonumber \\
&&\exp\left(-{1 \over 2}
\left\{(\hat{\bi{s}}+\delta \bi{s})^{\dag} \bi{S}^{-1}(\hat{\bi{s}}+\delta \bi{s})+\left[\bi{d}-\bi{f}(\hat{\bi{s}}+\delta \bi{s})\right]^{\dag} \bi{N}^{-1}\left[\bi{d}-\bi{f}(\hat{\bi{s}}+\delta \bi{s})\right]\right\}\right).
\label{lik1}
\end{eqnarray}
If we collect all the terms that involve $\delta \bi{s}$ they will vanish by the requirement that the gradient is zero, equation \ref{gder1}, by definition 
of $\hat{\bi{s}}$: the posterior surface is quadratic at the minimum of the 
loss function. We are left with
\begin{eqnarray}
&&P(\bi{s},\bi{d} \vert \bi{\Theta})= (2\pi)^{-(M+N)/2} \det(\bi{SN})^{-1/2} \times \nonumber \\
&&\exp\left(-{1 \over 2}
\left\{\hat{\bi{s}}^{\dag} \bi{S}^{-1}\hat{\bi{s}}+\left[\bi{d}-\bi{f}(\hat{\bi{s}})\right]^{\dag} \bi{N}^{-1}\left[\bi{d}-\bi{f}(\hat{\bi{s}})\right]+\delta \bi{s}^{\dag}\bi{D}\delta \bi{s}\right\}\right) \times \nonumber \\
&&\exp
\left[ \bi{D}_3(\hat{\bi{s}})\delta \bi{s}\delta \bi{s}\delta \bi{s}+\bi{D}_4(\hat{\bi{s}})\delta \bi{s}\delta \bi{s}\delta \bi{s}\delta \bi{s}+...\right],
\label{lik2}
\end{eqnarray}
where $\bi{D}$ is the curvature matrix of equation \ref{D1} and more generally we have defined 
\begin{equation}
\bi{D}_n(\hat{\bi{s}})={1 \over 2n!}\left[{\partial^n \chi^2\over (\partial \bi{s})^n}\right]_{\bi{s}=\hat{\bi{s}}}. 
\end{equation}
We see that $\bi{D}=2\bi{D}_2$. 
In terms of derivatives of $\bi{f}$ we have 
\begin{equation}
\bi{D}_3={1 \over 4} \bi{f'}^{\dag}\bi{N}^{-1}\bi{f''},
\end{equation}
and 
\begin{equation}
\bi{D}_4={1 \over 48} \left[3\bi{f''}^{\dag}\bi{N}^{-1}\bi{f''}+4\bi{f'}^{\dag}\bi{N}^{-1}\bi{f'''}\right].
\label{D4}
\end{equation}
We have consistently dropped the terms with $\bi{d}-\bi{f}(\hat{\bi{s}})$, which fluctuate around zero. All the terms in the last line of equation 
\ref{lik2} are scalars. 

Next we assume a perturbative expansion in terms of higher order derivatives
and/or low noise (we will define the expansion parameter below), 
bringing the higher order terms $\bi{D}_n$ down from the exponential, 
\begin{equation}
\exp\left[\bi{D}_3(\hat{\bi{s}})\delta \bi{s}\delta \bi{s}\delta \bi{s}+\bi{D}_4(\hat{\bi{s}})\delta \bi{s}\delta \bi{s}\delta \bi{s}\delta \bi{s}+...\right]=1+\bi{D}_3(\hat{\bi{s}})\delta \bi{s}\delta \bi{s}\delta \bi{s}+\bi{D}_4(\hat{\bi{s}})\delta \bi{s}\delta \bi{s}\delta \bi{s}\delta \bi{s}+...
\end{equation}

We can now obtain the perturbative expansion of the marginalized likelihood function by integrating out $\delta \bi{s}$,
\begin{eqnarray}
&&P(\bi{\Theta} | \bi{d})  \propto  L(\bi{d}|\bi{\Theta})
=  \int P(\bi{s},\bi{d} \vert \bi{\Theta}) d^M\delta \bi{s}  \nonumber \\
&&=(2\pi)^{-(M+N)/2} \det(\bi{SN})^{-1/2}
\exp\left(-{1 \over 2}\left[\hat{\bi{s}}^{\dag}\bi{S}^{-1}\hat{\bi{s}}+(\bi{d}-\bi{f}(\hat{\bi{s}}))^{\dag}\bi{N}^{-1}(\bi{d}-\bi{f}(\hat{\bi{s}}))\right]\right)
 \times \nonumber
\\ &&\int  \exp\left\{-{1 \over 2}\delta\bi{s}^{\dag}
\bi{D}\delta\bi{s}\right\}d^M\bi{s} \left[ 1+\bi{D}_3(\hat{\bi{s}})\delta \bi{s}\delta \bi{s}\delta \bi{s}+\bi{D}_4(\hat{\bi{s}})\delta \bi{s}\delta \bi{s}\delta \bi{s}\delta \bi{s}...\right] \nonumber
 \\
&&= (2\pi)^{-N/2}\det(\bi{SN}\tilde{\bi{D}})^{-1/2} 
\exp\left(-{1 \over 2}\left[\hat{\bi{s}}^{\dag}\bi{S}^{-1}\hat{\bi{s}}+(\bi{d}-\bi{f}(\hat{\bi{s}}))^{\dag}\bi{N}^{-1}(\bi{d}-\bi{f}(\hat{\bi{s}}))\right]\right). 
\label{marg}
\end{eqnarray}
We have defined 
\begin{equation}
\det \tilde{\bi{D}}^{-1/2}=\det \bi{D}^{-1/2}\left[1+3{\rm tr}(\bi{D}^{-1}\bi{D}_4\bi{D}^{-1})+...\right].
\end{equation}
The lowest order term $\det \bi{D}$ is given by the gaussian integral. The next 
order term is a 1-loop contribution consisting of $f''^2$ and $f'''f'$
inside $\bi{D}_4$, 
and we used Wick's theorem to perform the gaussian
integrals. 
We can also define
\begin{equation}
\tilde{\bi{C}} \equiv \bi{SN\tilde{D}} = \bi{N}+\bi{f'}\bi{S}\bi{f'}^{\dagger}+...,
\label{C}
\end{equation} 
where $...$ denotes higher order terms in $\tilde{\bi{D}}$ and 
we used equation \ref{D} for $\bi{D}$. 

We see that to marginalize over a set of parameters $\bi{s}$ one has first to
find  their maximum posterior $\hat{\bi{s}}$, as derived in previous section via optimization,
after which the integration
over the parameters can be performed analytically assuming a 
perturbative expansion. This procedure implicitly assumes a single maximum posterior and 
a simple posterior around $\hat{\bi{s}}$. When this is 
not satisfied, for example in the presence of multiple 
peaked posterior, one must either 
perform the appropriate statistical average over these, if they 
are widely separated, or show that all the local peaks are close to 
each other and thus essentially 
giving the same solution. 
For now we will proceed assuming a single peaked
posterior. Note that integration over $\bi{s}$ produced an additional 
factor of $\tilde{\bi{D}}^{-1/2}$ relative to doing the maximum likelihood over all parameters simulteneously.

Let us look at the structure of the resulting expression in equation \ref{marg}. The main term 
that depends on the data is in the exponential, $\exp[-\chi^2(\hat{\bi{s}})/2]$. This term consists 
of the prior term and of the goodness of data fit, both evaluated
at the MAP position $\hat{\bi{s}}$. The other term is $\det \bi{SN}\tilde{\bi{D}}$. This term represents the volume of the 
priors and posteriors. We wish to explore its sensitivity to the 
parameters $\bi{\Theta}$ that determine $\bi{S}$, so we wish to 
take derivative of this term wrt $\bi{\Theta}$. 

From equation \ref{C} we obtain
\begin{equation}
\det(\tilde{\bi{C}}) \approx
\det(\bi{N}+\bi{f'} \bi{S}\bi{f'}^{\dag}).
\end{equation}
In the low noise limit it gives $\det(\tilde{\bi{C}}) \approx \det (\bi{f'} \bi{S}\bi{f'}^{\dag})$, which is highly sensitive to the 
parameters $\bi{\Theta}$ inside $\bi{S}$. In the high noise limit we have $\det(\bi{SND}) \approx \det\bi{N}$. In this limit we do not get any information about the parameters $\Theta$. This
is reflected in the values of $\hat{\bi{s}}$, which are at the their full value in the low noise limit and become 
smaller and smaller as the noise increases. 

Next we look at the data dependence of the curvature matrix $\tilde{\bi{D}}$. 
If the model is linear $\tilde{\bi{D}}$ does not 
depend on the data, because $\bi{f'}$ is a constant and 
$f^{(n)}=0$ for $n>1$. Thus this term can be evaluated using 
a method that is data independent, such as a simulation. 
For nonlinear models we have to account for the data dependence of 
$\tilde{\bi{D}}$. 
Let us first look at the leading term $\bi{D}$, 
given in equation \ref{D}. For nonlinear models the first derivative 
$\bi{f'}(\hat{\bi{s}})$ 
is data dependent. For example, 
suppose we Taylor expand the forward model in terms of $\bi{s}$,
\begin{equation}
\bi{f}(\bi{s})=\bi{R}_1
\bi{s}+{1 \over 2}\bi{R}_2\bi{s}\bi{s}...,
\end{equation}
we see that 
\begin{equation}
\bi{f'}(\hat{\bi{s}})=\bi{R}_1+\bi{R}_2\hat{\bi{s}}+...
\end{equation}
and hence the covariance $\bi{D}$ depends on $\hat{\bi{s}}$ 
through $\bi{R}_2\hat{\bi{s}}+...$. Depending on the realization value of 
$\hat{\bi{s}}$ the associated error can be larger or smaller than 
the first order term $\bi{R}_1$. 
We can see this from equation \ref{gder1}. The solution to this 
equation balances between the prior term $\bi{S}^{-1}\hat{\bi{s}}$, 
which on its own would give $\hat{\bi{s}}=0$, and the data term
$\bi{f'^{\dag}(\hat{s}){N}^{-1}[d-f(\hat{s})}]$, which on its own would 
solve for $\bi{f}(\hat{\bi{s}})=\bi{d}$. The balance between the two 
is determined by $\bi{Sf'^{\dag}(\hat{s}){N}^{-1}}$, which contains 
data realization dependence inside $\bi{f'}(\hat{\bi{s}})$. 
Thus the solution for 
$\hat{\bi{s}}$ automatically accounts for generalized noise 
weighting of the modes including realization dependence, 
at least at the lowest order. 

Let us look next at the 1-loop term ${\rm tr}(\bi{D}^{-1}\bi{D}_4\bi{D}^{-1})$. This term represents an increase in volume of posterior of $\delta \bi{s}$ 
around $\hat{\bi{s}}$ relative to the gaussian approximation. 
Using equations \ref{D} and \ref{D4}
we find this term scales as, schematically,
\begin{equation}
\bi{D}^{-1}\bi{D}_4\bi{D}^{-1} \propto [3\bi{f''}^2+4\bi{f'f'''}]\bi{N}. 
\end{equation}
Thus the 1-loop term vanishes in the limit of either small noise or 
small nonlinearity. Both of these conditions are likely to be satisfied 
on large scales, where noise relative to power is small and modes are 
nearly linear. As we push the marginalization to smaller scales we get larger 
and larger corrections from this term. This term is also data 
dependent. This means that one may have a situation where Taylor 
expansion suggests the posterior is wide (narrow), because the variance 
of the mode $\bi{D}^{1/2}$ is large (small), but the full posterior may be narrower (wider) once we include higher order terms, such as curtosis
computed here. At this order our procedure will cease to be optimal
in terms of optimal mode weighting, but we will still make it 
unbiased, as described below. It is unclear however how suboptimal due 
to higher order corrections our method is. We will show below that 
our method is essentially giving equal weight to all reconstructed modes, 
and this may well be valid even beyond the formal applicability regime 
derived here. 

Optimization methods in large number of 
dimensions do not provide $\det \bi{D}$, although 
approximations to it exist in certain methods (e.g. BFGS). 
It seems even less promising to be 
able to compute analytically $\det \tilde{\bi{D}}$. 
While 
this term is data realization dependent, it is computed independently 
of the $\exp[-\chi^2(\hat{\bi{s}})]$ term, and is essentially 
determined by the 
posterior volume of $\delta \bi{s}$ 
around $\hat{\bi{s}}$. 
In a large number of dimensions the average of 
this term will not be strongly data realization 
dependent: for any given realization we will have about the same 
level of mode to mode fluctuations that will on average have the same 
posterior volume and give the same value of 
$\det \tilde{\bi{D}}$ regardless of where $\hat{\bi{s}}$ is. 
This suggests we may evaluate this term average using Monte Carlo methods, with 
a realistic simulation of the data: even 
though the MAP values $\hat{\bi{s}}$ in a simulation will have no 
relation to the corresponding MAP of the data, 
the posterior volume of $\delta \bi{s}$ 
around it will be nearly the same as for the data, as long as the 
simulation is close enough to the data. 
We may give up some 
optimality by not properly weighting the modes by their full posterior volume, but our procedure is exact in the linear case and in the low 
noise case, and as argued above equal weighting of the modes may be the 
valid approach even beyond these two formal limits. In any case, 
we cannot evaluate 
the posterior volume properly using the methods developed here, which are perturbative anyways and so not valid in general. This will be the basis for 
our efficient method of evaluating the optimal estimator and its 
covariance matrix. Note that even though we use random simulations in 
this paper to compute the volume term, we may also 
construct a simulation that is close to the actual data realization, 
further reducing the dependence of the volume term on the data realization. 

\subsection{Maximizing the bandpower likelihood}

Using the Bayes theorem and assuming a flat prior on $\bi{\Theta}$ we have
interpreted the posterior 
$P(\bi{\Theta}|\bi{d})$
as proportional to the 
 likelihood $L(\bi{d}|\bi{\Theta})$. 
We can justify using flat prior as the choice which gives unbiased estimators of cosmological parameters,
when the summary statistics $\bi{\hat{\Theta}}$ are used 
to determine them. This interpretation is only valid when the 
noise bias and Fisher matrix are evaluated using a fiducial power 
spectrum not affected by the measurements. When the
fiducial model is too far from the final solution one can repeat 
the procedure with a fiducial model that is closer to the actual 
solution. Note that we do not impose positivity of the bandpowers
with the prior: this is justified if the bandpowers are viewed 
as summary statistics that are subsequently used to determine 
cosmological parameters (where physical priors can be imposed). 

The next step is to maximize this  likelihood with respect to the bandpowers $\bi{\Theta}$. Given a set of measurements $\bi{d}$ we wish to find the most probable
set of bandpowers $\bi{\Theta}$ from equation \ref{marg},  
where $\bi{D}$, $\bi{S}$ and $\hat{\bi{s}}$ implicitly depend on the bandpowers $\bi{\Theta}$.
To find the most probable set of parameters one needs to find the
maximum of the  likelihood function $L(\bi{\Theta})$. 

We will find the maximum posterior using Newton's method. 
We wish to expand the log  likelihood in terms of $\bi{\Theta}$ to a quadratic order around some fiducial values,
\begin{equation}
\ln L(\bi{\Theta}_{\rm fid}+\Delta \bi{\Theta})=\ln L(\bi{\Theta}_{\rm fid})+\sum_l \left[{\partial \ln L(\bi{\Theta}) \over \partial \Theta_l} \right]_{\bi{\Theta}_{\rm fid}}\Delta \Theta_l+ {1 \over 2} \sum_{ll'}\left[{\partial^2 \ln L(\bi{\Theta}) \over \partial \Theta_l \partial \Theta_{l'}}\right]_{\bi{\Theta}_{\rm fid}}\Delta \Theta_l\Delta \Theta_{l'},
\label{llik}
\end{equation}
where the terms are evaluated at the fiducial model $\bi{\Theta}_{\rm fid}$. 
We will seek a solution where the first derivative vanishes using Newton's
method. The last term of equation \ref{llik} defines the curvature matrix as the second 
derivatives of log likelihood with respect to the parameters. 

To get the first derivative we need to take a derivative of log  likelihood in equation \ref{marg} with respect to 
$\Theta_l$. The dependence on $\Theta_l$ is in $\hat{\bi{s}}$, $\bi{S}$ and $\bi{f}(\hat{\bi{s}})$. We have 
\begin{equation}
{\partial\left[(\bi{d}-\bi{f}(\hat{\bi{s}}))^{\dag}\bi{N}^{-1}(\bi{d}-\bi{f}(\hat{\bi{s}}))\right] \over \partial\Theta_l}=-2\left[\bi{R}{\partial\hat{\bi{s}} \over \partial\Theta_l}\right]^{\dag}\bi{N}^{-1}(\bi{d}-\bi{f}(\hat{\bi{s}}))=-2\left[{\partial\hat{\bi{s}} \over \partial\Theta_l}\right]^{\dag}\bi{S}_{\rm fid}^{-1}\hat{\bi{s}},
\end{equation}
where the first relation follows from equation \ref{rij} and the last relation follows from equation \ref{gder1}. We also have 
\begin{equation}
{\partial\left[ \hat{\bi{s}}^{\dag}\bi{S}_{\rm fid}^{-1}\hat{\bi{s}}\right]\over \partial\Theta_l}=2\left[{\partial\hat{\bi{s}} \over \partial\Theta_l}\right]^{\dag}\bi{S}_{\rm fid}^{-1}\hat{\bi{s}}-
2E_l(\bi{S}_{\rm fid},\hat{\bi{s}}),
\label{sss}
\end{equation}
defining
\begin{equation}
E_l(\bi{S}_{\rm fid},\hat{\bi{s}})={1 \over 2}\hat{\bi{s}}^{\dag}\bi{S}_{\rm fid}^{-1}\bi{\Pi}_l\bi{S}_{\rm fid}^{-1}\hat{\bi{s}}.
\label{el}
\end{equation}
All 
the matrix operations involve diagonal matrices, so using equation \ref{pilin} we get 
\begin{equation}
E_l(\bi{S}_{\rm fid},\hat{\bi{s}})={1 \over 2}\sum_{k_l}{\hat{s}_{k_l}^2 \over \Theta_{\rm{fid},l} S_{{\rm fid},k_l}},
\label{wfmv1}
\end{equation}
where the sum is over all modes within the bandpower and 
$S_{{\rm fid},k_l}$ is the fiducial 
power spectrum amplitude at mode $k_l$, which 
for narrow power spectrum bins is simply $\Theta_{{\rm fid},l}$.
We see that the terms with $d\hat{\bi{s}}/d\Theta_l$ cancel out, which 
gives us, using equation \ref{marg} 
\begin{equation}
{\partial \ln L(\bi{\Theta}) \over \partial \Theta_l}=E_l
-b_l ,
\label{like_deriv}
\end{equation}
where we defined 
\begin{equation}
b_l = {1 \over 2}{\rm tr}\left[{\partial \ln (\bi{SN\tilde{D}}) \over \partial \Theta_l}\right]_{\bi{S}_{\rm fid}}.
\label{bl}
\end{equation}
Note that $\bi{N}$ has no dependence on $\bi{\Theta}$. 
This term is non-zero 
because squaring the MAP field (as done in first term of rhs in equation \ref{like_deriv}) creates a bias.
In line with the linear 
analysis we will continue to call this term noise bias, but unlike the linear case this term 
may depend on both the noise and the signal. 
The first term in equation \ref{like_deriv} 
states that for narrow bins the derivative is given by adding up all the squares of the modes. We can see that in this term the 
curvature matrix $\bi{D}$ (or its generalization $\tilde{\bi{D}}$) does not enter, and its evaluation 
is not required. Since we cannot compute the curvature matrix we 
also cannot take its derivative with respect to parameters, which 
is needed to compute the noise bias in equation \ref{bl}. In this paper we instead compute 
the noise bias by evaluating its ensemble average, using a method described further below.
This will mean that we have given up on the information contained in this 
term. We expect this information to be subdominant, at least in the 
linear and low noise limits.
 
The maximum likelihood solution for $\hat{\bi{\Theta}}$ 
is given by 
\begin{equation}
\left[{\partial \ln L(\bi{\Theta}) \over \partial \Theta_l}\right]_{\hat{\bi{\Theta}}}=E_l(\hat{\bi{\Theta}})
-b_l(\hat{\bi{\Theta}})=0. 
\label{like_deriv1}
\end{equation}

To evaluate we would need to evaluate equation \ref{bl}. This gives
\begin{equation}
b_l = {1 \over 2}{\rm tr}\left[{\partial \ln \tilde{\bi{C}} \over \partial \Theta_l}\right]= {1 \over 2}{\rm tr}\left[{\partial \tilde{\bi{C}} \over \partial \Theta_l}\tilde{\bi{C}}^{-1}\right].
\label{blc}
\end{equation}
Using equation \ref{C} and 
dropping higher order terms we obtain
\begin{equation}
b_l={1 \over 2}{\rm tr}\left[\bi{f'}\bi{\Pi}_l\bi{f'}^{\dagger}\tilde{\bi{C}}^{-1}\right].
\end{equation}
We wish to solve equation \ref{like_deriv1} to obtain ML estimate 
of $\hat{\Theta}_l$, which appears inside $\bi{S}$. This is a 
complicated nonlinear equation in terms of $\bi{S}$ and hence 
one cannot write the solution in a closed form. 

Instead of solving this equation directly we will adopt Newton's method, 
where we use the quadratic expansion of log-likelihood in equation 
\ref{llik}, evaluate all the elements at $\bi{\Theta}_{\rm fid}$
where the log-likelihood derivative is not zero, and then use Newton's method 
to find the values  $\hat{\bi{\Theta}}$ where the first derivative 
is zero. To do this we must therefore evaluate the curvature matrix.
Instead of the actual curvature matrix for parameters we will often work with the Fisher matrix, defined as the ensemble average of the curvature around the maximum,
\begin{equation}
F_{ll'}=-\left\langle {\partial^2 \ln L(\bi{\Theta}) \over \partial \Theta_l \partial \Theta_{l'}} \right\rangle .
\label{fish1}
\end{equation}
Brackets denote ensemble averaging. It may be possible to compute the actual curvature matrix for the 
data (discussed further below and in appendix \ref{appg}), but in this paper we will not distinguish between 
the two and simply call it Fisher matrix, even though this is not 
exactly the same as the curvature matrix in the 
Bayesian analysis.
For modes that are gaussian 
distributed one can interpret the inverse of the Fisher matrix $\bi{F}^{-1}$ as an
estimate of the covariance matrix of the parameters $\hat{\bi{\Theta}} $,
\begin{equation}
\langle \Delta \hat{\bi{\Theta}}\Delta \hat{\bi{\Theta}}^{\dag}\rangle -
\langle \Delta \hat{\bi{\Theta}}\rangle\langle \Delta \hat{\bi{\Theta}}\rangle^{\dag}
=\bi{F}^{-1}.
\label{minvar}
\end{equation}
Note that for linear modes the 
likelihood function can be approximated as 
gaussian around the maximum provided that sufficient number of  independent 
modes contribute to each $\Theta_l$, by central limit theorem.
Appendix \ref{appa} discusses an approximation that goes beyond this limit. 

We are finally in position to use Newton's method to write the maximum likelihood estimator for $\Delta \bi{\Theta}$. 
The solution to the quadratic log likelihood at the peak can be 
obtained by setting 
the derivative of equation \ref{llik} with respect to $\bi{\Delta \Theta}$ to zero,
which upon inserting equation \ref{like_deriv} gives 
\begin{equation}
(\bi{F}\Delta \hat{\bi{\Theta}})_l=
E_l
-b_{l},
\label{wfmv}
\end{equation}
where $b_l$ is the noise bias contribution. To solve this equation 
one needs both $b_l$ and $F_{ll'}$. In principle we could get the 
latter by evaluating equation \ref{fish1}. This would give an expression 
in terms of a matrix multiplications and trace, 
\begin{equation}
F_{ll'}=
{1 \over 2}{\rm tr}\left[{\partial \tilde{\bi{C}} \over \partial \Theta_l}\tilde{\bi{C}}^{-1}{\partial \tilde{\bi{C}} \over \partial \Theta_{l'}}\tilde{\bi{C}}^{-1}
\right],
\label{fll3}
\end{equation}
which is formally of order 
$N^3$, if we had the matrices. But we do not actually have the matrix 
$\tilde{\bi{C}}$ or $\tilde{\bi{D}}$, so we cannot evaluate this even if we wanted to. We will 
thus evaluate the Fisher matrix using a different approach using 
ensemble averaging of simulations. However, 
in the special case of no mode coupling the Fisher matrix is 
diagonal and we can combine equations
\ref{like_deriv1}, \ref{blc} and \ref{fll3} to obtain
\begin{equation}
F_{ll'}={2E_l^2 \over K_l} \delta_{ll'}.
\label{fllel2}
\end{equation}
In this case we get the actual curvature matrix from the data itself. 
We will test the accuracy of this approximation in section \ref{sec5}
for the specific case of periodic boundary conditions, which is the 
only case where the assumption of uncorrelated modes may be valid. 

From equation \ref{wfmv} we see that the raw power spectrum $E_l$ of the 
reconstructed field requires noise bias subtraction. 
We also see that after noise bias subtraction we obtain the bandpower 
estimates convolved with the Fisher matrix $\bi{F}\Delta \hat{\bi{\Theta}}$. Fisher matrix thus 
describes both the covariance matrix and the bandpower mixing. It is the 
latter interpretation that will be the basis for our fast Fisher matrix 
evaluation method. As promised, equation \ref{wfmv1} agrees with our heuristic derivation in equation \ref{wfmv0}. The procedure is implied to 
be iterative: if the chosen fiducial model is not sufficiently close 
to the final answer one should choose a new fiducial model and repeat 
the procedure. The fiducial model should not be simply chosen to be the 
observed one given by 
$\bi{\Theta}_{\rm fid}+\Delta \hat{\bi{\Theta}}$, since that contains noise and sampling variance fluctuations. Instead, 
it should be a smooth model that is sufficiently close to it. 
Equation \ref{wfmv} suggests that all the information from the 
data is inside $E(\bi{S}_{\rm fid},\hat{\bi{s}})$. In reality, $\bi{b}$ and $\bi{F}$
may also be data dependent (although this is explicitly not the case for 
the linear model). This is discussed further below. 

The form given in equation \ref{wfmv} is not the only 
possibility, and may not be the most practical. More 
generally, one can write 
\begin{equation}
\Delta \hat{\Theta}_k^, =\sum_l M_{kl}\left[E_l-b_{l} \right],\, {\rm or\, in\,  matrix\, form}, \, \Delta\hat{\bi{\Theta}}^, =\bi{M}(\bi{E}-\bi{b}),
\label{hatth}
\end{equation}
where $\bi{M}$ is a matrix that can be chosen depending on the form in 
which we wish to provide the estimator $\Delta \bi{\Theta}^,$, 
which can be 
viewed as a window convolved version of $\Delta \bi{\Theta}$. 
There are three natural choices \cite{2003NewA....8..581P} 
that we discuss next. 

First choice is 
$\bi{M}=\bi{F}^{-1}$, which corresponds to the fully deconvolved estimator ($\Delta \bi{\Theta}^,=\Delta \bi{\Theta}$ from equation \ref{wfmv}), with anti-correlated errorbars (which can explode if the sampling of modes 
is too fine compared to the width of the window). 

A second choice is 
\begin{equation}
M_{kl}={\delta_{kl} \over \sum_i F_{ki}},
\label{mmin}
\end{equation}
which corresponds to the minimum variance errors on 
$\Delta \hat{\Theta}^,_k$, but which are all convolved with a window
given by $\bi{MF}$. We will use this choice in this paper because it does not require a
matrix inversion.
Division by $\sum_i F_{ki}$ ensures the estimator is 
centered around the true value. However, this can cause problems if the matrix $\bi{M}$ is singular, which can happen if there is 
no information about a given bandpower in the data. In this 
case it is better not to divide by this factor and simply 
use $M_{kl}=\delta_{kl}$.

A third choice is to decorrelate the estimators (within the disconnected covariance matrix approximation) \cite{1997MNRAS.289..295H},
\begin{equation}
M_{kl}={F^{-1/2}_{kl} \over \sum_i F_{ki}^{1/2}}.
\label{mdec}
\end{equation}

With any of these definitions we can write the expectation value 
and variance of the estimator as 
\begin{eqnarray}
\langle \Delta\hat{\bi{\Theta}}^, \rangle &=&\bi{MF} \Delta \bi{\Theta} \equiv \bi{W \Delta \Theta}, \label{wind} \\
\langle \Delta\hat{\bi{\Theta}}^, \Delta \hat{\bi{\Theta}}^{,\dag} \rangle &-&\langle \Delta \hat{\bi{\Theta}}^, \rangle \langle \Delta \hat{\bi{\Theta}}^{,\dag} \rangle=\bi{MFM} \equiv \bi{C},
\label{covfish}
\end{eqnarray}
where the latter contains only the connected part of covariance matrix
and we defined the window matrix $\bi{W}=\bi{MF}$ and the covariance 
matrix $\bi{C}=\bi{MFM}$.
All of these choices are equivalent in terms of their information content: 
there is no gain or loss of information in choosing one over the other. All the data 
compression has been done in equation \ref{hatth}. The first choice
corresponds to identity window matrix, the second is minimum variance 
and does not require any matrix
inversion (equation \ref{wind}) 
and is thus the easiest to compute, while the third 
choice provides a diagonal covariance matrix of the estimator, so bandpowers 
are easy to combine. 

The estimator is unbiased as long as the Fisher matrix is computed 
using a fiducial power spectrum, not affected by the data. If the 
chosen fiducial power spectrum is very different from the one 
preferred by the data the procedure is still unbiased, but the estimator
may not be optimal and the Fisher matrix may not be accurately interpreted
as a covariance matrix. In these situations one may repeat the analysis 
with a chosen fiducial power spectrum closer to the one preferred by the
data. 
In the case of high precision cosmology where the deviations between the 
true and fiducial power spectrum are small, the corresponding error induced on the 
covariance matrix also becomes small. 

The estimator presented here (quadratic in MAP) approximates the posterior for 
bandpowers as a multi-variate gaussian. This allows us to normalize the 
posterior, and assuming $L$ bandpowers we have 
\begin{equation}
P(\Delta\bi{\Theta} | \bi{d}) =[(2\pi)^{L} \det\bi{C}]^{-1/2} 
\exp\left(-{1 \over 2}
\left[\bi{W}\Delta\bi{\Theta}-\Delta\hat{\bi{\Theta}}^,\right]^{\dag} \bi{C}^{-1}\left[\bi{W}\Delta\bi{\Theta}-\Delta\hat{\bi{\Theta}}^,\right]\right).
\end{equation}

In the linear case this is valid in the limit of large number of modes by central limit theorem. Typically one combines different bandpowers 
to determine a small number of cosmological paramaters, which makes 
central limit theorem even more valid. However, in special cases, such as 
clustering on very large scales with few measured modes, the small number of modes 
requires one to use a more accurate posterior probability distribution. We discuss 
this situation in appendix \ref{appa}.
The connected part of the 
covariance matrix is not given by the 
Fisher matrix construction presented above. 
However, since we are estimating the linear 
power spectrum, which is assumed to be gaussian, 
it is likely that there is no significant connected 
part, at least for low noise. To address this properly 
one needs to compare the disconnected covariance 
matrix to the one from a large number of mocks where 
the whole procedure has been simulated, and we will 
not investigate this in any detail in this paper. 

\subsection{An efficient evaluation of noise bias and Fisher matrix}

What is left is to evaluate the noise bias and Fisher matrix. 
This is often the most expensive part of the calculation, even for the linear model \cite{2003NewA....8..581P}. 
Here we are solving the nonlinear problem in which the 
minimization cannot be solved with linear algebra anyways, so we have to devise a new scheme to evaluate these terms. 

Before proceeding it is useful to clarify the nature of the whole procedure. We began by estimating the modes that minimize the variance of the data given the noise and the prior. In doing so 
we have chosen a fiducial power spectrum $\bi{S}_{\rm fid}$ for the prior. 
We then square these modes and add them up within the bandpower $l$. We
want to make this estimator unbiased, 
and with the correct window function describing correlations between the bandpowers. Our primary task is to have the estimator that is unbiased for any choice of the fiducial power 
spectrum $\bi{\Theta}_{\rm fid}$. In practice, we expect 
the estimator will be almost as good as minimum variance even if the fiducial power 
spectrum chosen 
in the minimization procedure is not exactly the true power spectrum.
Fisher matrix plays two distinct roles in the parameter estimation, as a window function (equation \ref{wind}) and as a covariance matrix (equation \ref{covfish}). 
Here we will use its interpretation as the window function to derive it,
quantifying the 
sensitivity of one bandpower to another. 
This only involves two point functions, 
and does not rely on the assumption of gaussianity. But Newton's 
method guarantees that the resuting answer is also the 
covariance matrix. 

Since a direct evaluation is not computationally feasible, we will be evaluating Fisher 
matrix and noise bias terms using simulations. The first question is whether these terms 
depend on the actual realization. 
For example, the noise bias is 
given by the derivative of the log determinant of the curvature matrix with respect to the 
bandpowers (equation \ref{bl}) and this can depend on the actual realization.
This is however not the case in the linear limit \cite{1998ApJ...503..492S}. It is also not very important 
in the low noise limit, where most of the information is in the reconstructed modes
and the noise bias is small. More generally, we have argued in the 
beginning of this section that the volume element $\det \tilde{\bi{D}}$ 
of $\delta \bi{s}$ around $\hat{\bi{s}}$ 
can 
be determined from a simulation in the limit of large number of modes
even if the values of $\hat{\bi{s}}$ differ between a simulation and 
the data. We will thus proceed by assuming that there is no 
realization dependence of Fisher matrix and noise bias, and we will 
determine them 
using simulations that are unrelated to the actual realization given by the data. 
Should this assumption turn out to be violated one could explore realization dependence
of these terms, for example by using realizations that are conditioned 
on the data (such as Gibbs sampling), but we do not pursue this further in this paper. 
We emphasize that our estimator will still be unbiased at the fiducial model and small
variations around it, but it may not be minimum variance. 

For the simulations we thus first generate a gaussian random realization of the signal in Fourier space $\bi{s}_s$ at the fiducial model, such that
\begin{equation}
\langle \vert \bi{s}_s \vert^2 \rangle=\bi{S}_{\rm fid}.
\label{dssim}
\end{equation}
We have explored both a gaussian realization, which contains 
sampling variance fluctuations, 
and the fixed norm of the mode, where only 
the phase of the mode is random, while 
the mode square is exactly given by the fiducial power. 
The latter does not suffer 
from sampling variance and converges faster if no sampling variance 
correction is used, although we will also give prescriptions how to handle 
sampling variance for the former case. Suppression of sampling variance is 
crucial and guarantees faster convergence, so the procedure here should 
not be viewed simply as a Monte Carlo simulation approach. 

Next, we map it into the observational space using a forward model,
\begin{equation}
\bi{d}_s=f(\bi{s}_s).
\end{equation}
We also generate a noise sample. 
Similar to the signal, 
we have explored both a gaussian realization with unit variance, 
and the real stochastic $Z_2$ estimator, where for each real space 
point $\bi{r}_i$ we generate a random number consisting of 1’s and -1’s. 
The latter has a fixed amount of power and is expected to converge faster. 
In both cases 
we multiply with $\sigma(\bi{r}_i)$, the square root of the noise variance at that point $N(\bi{r}_i)=\sigma^2(\bi{r}_i)$, to create a random noise 
data vector $\bi{d}_n$ with the correct variance. 
We use this noise 
realizations $\bi{d}_n$ added to the signal data $\bi{d_s}$ to create
\begin{equation}
\bi{d}_{s+n}=\bi{d}_s+\bi{d}_n.
\label{dsim}
\end{equation}
Now that we have a data simulation 
we pass this through 
the same sequence of optimization steps 
as the real data. 
This leads us to 
the simulated realizations of the mode estimator $\hat{\bi{s}}_{s+n}$
for the fiducial model. The cost for performing a single 
realization is comparable to the cost of analyzing the data. 

We want to ensure that the final estimates are unbiased.
From equation \ref{wfmv} the noise 
bias is 
\begin{equation}
b_l= 
E_l(\bi{\Theta}_{\rm fid},\hat{\bi{s}}_{s+n}). 
\label{blmv}
\end{equation}
This equation is simply stating that for a fiducial 
simulation the estimate of bandpower relative to 
fiducial power has to vanish. If needed, this needs to 
be averaged over several realizations.

The window function interpretation of the Fisher matrix is 
a response of a bandpower $l$ to another bandpower at $l'$, as in equation 
\ref{wfmv}.
A way to evaluate it is to inject power into one bandpower and 
measure the estimator response in all the bandpowers (in statistics this is called 
a sensitivity analysis). 
Specifically, 
we take the fiducial power spectrum realization from equation \ref{dsim}. 
We create another gaussian realization as a small perturbation to it, 
by injecting a small amount of power into the modes at a single
bandpower $l'$. 
For each of the modes $s_s(k_{l'})$ 
within the bandpower $l'$ 
we have a choice to generate either a random realization of the phase, 
or one given by $s_s(k_{l'})$. 
We add it to fiducial realization $\bi{s}_s$
\begin{equation}
\bi{s}_{l',s}=\bi{s}_s+\Delta \bi{s}_{l'}.
\label{sl}
\end{equation}
We want 
the injected power to be 
$\Delta \Theta_{l'}\bi{\Pi}_{l'}$ 
(where  $\Delta \Theta_{l'}$ should be small so that linearization of the Fisher matrix is valid, but not too 
small to be susceptible to roundoff errors), so we choose $\Delta \bi{s}_{l'}$
to satisfy
\begin{equation}
\langle \vert \Delta \bi{s}_{l',s} \vert^2 \rangle-
\langle \vert \Delta \bi{s}_{s} \vert^2 \rangle
=\Delta \Theta_{l'}\bi{\Pi}_{l'}.
\end{equation}
We have found that the injected power with modes in phase with $\bi{s}_s$
works better in terms of the noise in the Fisher matrix. 

We then run a forward model to generate $\bi{d}_{l',s}=f(\bi{s}_{l',s})$, 
to which we add exactly the same noise realization as the one in 
equation \ref{dsim}, $\bi{d}_{l',s+n}=\bi{d}_{l',s}+\bi{d}_n$. 
We pass it through 
the same sequence of optimization steps
as 
the fiducial model
simulation. Since we already have 
the optimization solution for the fiducial model simulation at 
$\bi{d}_s+\bi{d}_n$, 
and the injected modes $\Delta \bi{s}_{l'}$ are 
a small perturbation in a single bandpower, we can 
start the optimization at $\hat{s}_{s+n}$. 
This requires very few steps to converge, giving 
$\hat{\bi{s}}_{l',s+n}$. We compute for each bandpower 
\begin{equation}
\Delta \hat{\bi{s}}_{l',s+n}=\hat{\bi{s}}_{l',s+n}-\hat{\bi{s}}_{s+n}.
\end{equation}

We can apply equation \ref{wfmv} to $\Theta_{l'}$ and $\Theta_{l'}+\Delta 
\Theta_{l'}$, but still performing the optimization around the fiducial 
model, 
\begin{eqnarray}
(\bi{F}\bi{\Theta}_{\rm fid})_l&=&
E_l(\bi{\Theta}_{\rm fid},\hat{\bi{s}}_{s+n})
-b_{l}(\bi{\Theta}_{\rm fid},\hat{\bi{s}}_{s+n}), \nonumber \\
(\bi{F}[\bi{\Theta}_{\rm fid}+\Delta 
\bi{\Theta}_{l'}])_l&=&
E_l(\bi{\Theta}_{\rm fid},\hat{\bi{s}}_{l',s+n})
-b_{l}(\bi{\Theta}_{\rm fid},\hat{\bi{s}}_{l',s+n}).
\label{ff}
\end{eqnarray}
We are assuming that the Fisher matrix changes are at higher order, so that $\bi{F}$
is the same at the fiducial model and small perturbation away from it. 
In general, the noise bias depends on the signal and we cannot 
assume $b_{l}(\bi{\Theta}_{\rm fid},\hat{\bi{s}}_{s+n})=b_{l}(\bi{\Theta}_{\rm fid},\hat{\bi{s}}_{l',s+n})$. 
However, in the linear regime this condition is 
satisfied, and even when it is not we still obtain an estimator that is unbiased, 
but whose covariance may not be given by the resulting $\bi{F}$. 

Taking the difference between the two terms in equation \ref{ff} 
we obtain
\begin{equation}
\left[F_{ll'}+ {\partial b_l \over \partial \hat{\bi{s}}} { \partial \hat{\bi{s}}\over \partial \Theta_{l'}}\right]\Delta \Theta_{l'}\equiv 
\tilde{F}_{ll'}\Delta \Theta_{l'}=
E_l(\bi{\Theta}_{\rm fid},\hat{\bi{s}}_{l',s+n})-E_l(\bi{\Theta}_{\rm fid},\hat{\bi{s}}_{s+n})=
E_l(\bi{\Theta}_{\rm fid},\Delta\hat{\bi{s}}_{l',s+n}),
\label{fisher}
\end{equation}
where the last equality follows if the additional injected modes are uncorrelated with fiducial modes. 
We have defined a new matrix $\tilde{\bi{F}}$ 
as the sum of the true Fisher matrix and the matrix of derivatives of 
noise bias, which will only approximate the true Fisher matrix (although the relation is 
exact in linear case where the noise bias derivative term vanishes). 
In the remainder of this paper we will ignore this distinction between $\bi{F}$ and $\tilde{\bi{F}}$, although we will 
verify the quality of resulting 
covariance matrix in our tests. 
The Fisher matrix is then obtained from equation \ref{fisher},
averaged over several realizations if needed. It is important to recognize 
that our final expressions do not explicitly 
evaluate the curvature matrix $\bi{D}$, hence any optimization method can be used to find MAP. An alternative method using the auto and cross 
power of reconstructed modes is presented in appendix \ref{appg}.

The Fisher matrix is rapidly changing with the bandpower $l$, 
mostly because typically both the number of modes $K_l$ and the 
fiducial power $\Theta_l$ are rapidly changing with $l$. However, we can also 
define normalized version 
\begin{equation}
\langle T^2 \rangle_{ll'}=2\Theta_l \Theta_{l'}F_{ll'}/(K_lK_{l'})^{1/2}, 
\label{t2}
\end{equation}
which can be 
viewed as an average squared transfer function as in equation 
\ref{fisher_transfer}. This has the advantage of being a slowly 
changing function of $l$. As a consequence, one does not need to 
evaluate it at every $l$, especially for narrow bandpower bins. 
This allows 
one to determine the Fisher matrix by injecting the 
power into sparsely 
separated bandpowers, and then interpolate between averaged squared transfer function to get the full Fisher matrix. 

Sometimes we can adopt a simple 
approximation of equation \ref{fllel2}
(see appendix \ref{appg} and \cite{2003PhRvD..68h3002H}),
which follows from the covariance interpretation of the Fisher matrix, 
together with the gaussian variance approximation. 
We will show below this works remarkably well in simple periodic 
box test cases. 

\section{Beyond the bandpowers}
\label{sec4}

\subsection{Modeling the forward model parameters}

In the previous section we have assumed that the mapping between the 
initial modes, whose statistics is
described with the initial power spectrum $\bi{S}$, and the data is unique, in which 
case all the information is in the optimal power spectrum estimators of the 
bandpowers of $\bi{S}$. In reality this is almost never the case. First of all, 
the nonlinear mapping $\bi{f}(\bi{s})$ depends on other parameters that determine 
the growth of structure, such as matter density, dark energy equation of state
etc. Observations are also not directly in terms of matter over-density in 
comoving coordinates: 
for example, weak lensing observations can be expressed as 
an integral over the over-density times the mean matter density, and are 
typically observed as a function of angle, which can be converted to physical 
scale assuming cosmological model parameters such as dark matter, dark 
energy density etc. In addition, there are uncertainties in the 
physical modeling, such as baryonic effects, which depend on unknown 
astrophysical parameters and can be 
parametrized in terms of these. There are also data modeling uncertainties, 
for example, in the case of weak lensing 
both multiplicative and additive shear calibration 
biases and intrinsic alignments \cite{2004MNRAS.353..529H}. 

When it comes to galaxies 
the situation is even more complicated, since there are many ways to 
assign galaxies to dark matter halos, and one needs to allow for 
all models that are 
consistent with our present understanding of galaxy formation physics, yet currently cannot be determined from an ab initio modeling. For 
example, the simplest way is to parametrize galaxy clustering in terms of a 
free galaxy bias, which cannot be assumed to be known, but 
needs to be marginalized over.  
In addition, there are 
forward model parameters that control redshift space distortions (growth of structure) and mapping 
from comoving distances to angular and radial distance.

If the nuisance parameters affect only a few modes one can simply 
include these during the optimization process. For example, one may know 
the spatial templates that describe the systematics such as star contamination 
or dust extinction effects. The procedure to handle this situation 
is described in appendix \ref{appb}. 
In the absence of  
a prior the procedure is equivalent to setting the noise in these modes 
to a very 
large value and optimization sends the modes 
to zero. This eliminates cosmological information contained in those modes. 
For a few modes only 
this may not lead to a major loss of information. However, most of 
the nuisance parameters described above, such as galaxy bias, affect all 
the modes, and marginalizing over them during the optimization procedure
would send most of the modes to zero. Instead, 
these parameters must be handled 
at the level of  likelihood function analysis, similar to the 
power spectum bandpowers. 

Our general strategy how to handle these parameters will thus be 
the same as the 
bandpowers: we write the  likelihood function in terms of these
parameters and the bandpowers, expand it to second order, and use Newton's method to write 
down the solution at a quadratic order. 
Since these parameters enter 
in the nonlinear mapping $f(\bi{s})$ one must perform the corresponding 
partial derivatives. A very useful property of bandpowers is that they are 
linear in terms of covariance matrix $\bi{S}$, and as a result a single 
step Newton's method usually suffices, and their posterior is well described 
as a multivariate gaussian (or inverse-Wishart if the bandpower 
consists of a few modes only). 
This can no longer be assumed for
the parameters that affect the forward model, and one may need to iterate several times before reaching 
the maximum of the  likelihood. The posterior may also be more 
complicated than a gaussian (gaussian posterior is 
the reason that bandpowers
are preferred over the cosmological parameters as the expansion 
basis for $\bi{S}$, 
even though the number of parameters for the latter may be a lot lower). 
Here we will expand to quadratic order and leave the question 
of more accurate description of resulting posterior to future work. 

To proceed we will linearize forward model response around 
MAP solution $\hat{\bi{s}}$.
The  likelihood is given by 
equation \ref{marg}. We want to expand it in terms of the
forward model parameter $\lambda_l$. The terms 
containing derivative of $\hat{\bi{s}}$ cancel out as in equation 
\ref{like_deriv}. Let us look at the remaining 
data dependent term of the  likelihood that depends on $\lambda_l$, which gives
\begin{equation}
{\partial \ln L(\bi{\Theta},\bi{\lambda}) \over \partial \lambda_l}=
G_l
-\bi{b}_l ,
\label{like_deriv2}
\end{equation}
where
\begin{equation}
G_l(\bi{\lambda}_{\rm fid},\hat{\bi{s}})
=\left[{\partial \bi{f}(\hat{\bi{s}}) \over \partial \lambda_l}\right]_{\hat{\bi{s}}}^{\dag}
\bi{N}^{-1}\left[\bi{d}-\bi{f}(\hat{\bi{s}})\right]
\label{gl}
\end{equation} 
and the derivative of $\bi{f}$ with respect to $\lambda_l$ is to be evaluated at fiducial $\bi{\lambda}_{\rm fid}$ and fixed $\hat{s}$. 
This term is straight-forward to evaluate with the forward model: one computes forward model around the fiducial value (for example at
$\lambda_{{\rm fid},l} 
+ \delta \lambda_l$) and take the difference to the fiducial model to obtain 
\begin{equation}
\left[{\partial \bi{f}(\hat{\bi{s}}) \over \partial \lambda_l}\right]_{\bi{\lambda}_{\rm fid},\hat{\bi{s}}}= {\bi{f}(\hat{\bi{s}},\lambda_{{\rm fid},l}+\delta \lambda_l)- \bi{f}(\hat{\bi{s}},\lambda_{{\rm fid},l}) \over \delta \lambda_l}.
\label{lambda_deriv}
\end{equation}
Note that this must be evaluated at the actual realization consistent with the data. 
In the presence of noise this could be problematic: one possibility is to 
average over 
realizations consistent with the data, or use analytic expressions. An example 
is given below in the context of primordial non-gaussianity. 
Assuming the noise matrix is diagonal the sum 
is over all the data pixels $N$, 
\begin{equation}
G_l(\hat{\bi{s}})=\sum_{i=1}^N {\left(\partial f_i(\hat{\bi{s}})/ \partial \lambda_l\right)_{\hat{\bi{s}}}\left[d_i-f_i(\hat{\bi{s}})\right] \over N_i}.
\label{g2}
\end{equation}
This expression looks deceptively simple, but note that enumerator and 
denominator both vanish in the zero noise limit, so evaluating this 
expression can be numerically challenging in the low noise regime. We will 
convert it into mode space below. 
We also defined
\begin{equation}
\bi{b}_l = {1 \over 2}{\rm tr}[\partial \ln (\tilde{\bi{D}})/\partial \lambda_l] ,
\label{bl1}
\end{equation}
since matrix $\tilde{\bi{D}}$ is the only one that depends on the forward model parameters. 

From equation \ref{gl} it would appear that the
forward model parameters are limited by noise only, 
and not sampling variance. However, 
as already mentioned above the noise term $\bi{N}$ actually cancels out 
of this expression in the low noise limit. One can rewrite equation 
\ref{lambda_deriv} using equation \ref{gder1} into
\begin{equation}
G_l=\left[{\partial \bi{f}(\hat{\bi{s}}) \over \partial \lambda_l}\right]_{\hat{\bi{s}}}^{\dag} \bi{f'}^{-1}
\bi{S}^{-1}\hat{\bi{s}}.
\end{equation}

It is simplest to analyze this in 
the first order Taylor expansion case where $\bi{f}(\hat{\bi{s}})=\bi{f'}\hat{\bi{s}}$, 
in which case we obtain 
\begin{equation}
G_l=\hat{\bi{s}}{\partial \ln \bi{f'} \over \partial \lambda_l}
\bi{S}^{-1}\hat{\bi{s}}.
\label{glsl}
\end{equation}
Comparing to expression for $E_l$ in equation \ref{el} we see 
the two are the same if
\begin{equation}
\bi{\Pi}_l=2\bi{S}{\partial \ln \bi{ f'} \over \partial \lambda_l}.
\end{equation}
This is what one would expect if, for example, the forward model parameter
simply rescales the amplitude, in which case it is degenerate with the 
overall linear amplitude parameter $\bi{S} \propto \Theta_1^2$, for which $\bi{\Pi}_l=2\bi{S}/\Theta_1$. Another way to obtain the same result is 
by rewriting the data likelihood from 
equation \ref{marg} in the linearized case, using equations \ref{lwf} and \ref{D1} (dropping $\bi{f}"$ for
linearized models) into
\begin{equation}
L(\bi{d}|\bi{\Theta},\bi{\Lambda})=(2\pi)^{-N/2}\det(\tilde{\bi{C}})^{-1/2} 
\exp\left(-{1 \over 2}\bi{d}^{\dag}\tilde{\bi{C}}^{-1}\bi{d}\right),
\end{equation}
where $\tilde{C}$ is defined in equation 
\ref{C}. 
It is clear from this expression that the dependence of the likelihood 
with respect 
to the forward model parameters in $\bi{f'}$ and with 
respect to the power spectrum parameters in $\bi{S}$ appears 
in the same term $\bi{f'}\bi{S}\bi{f'}^{\dagger}$ contained inside $\tilde{\bi{C}}$. 

We can rewrite equation \ref{glsl} and make it applicable to the 
nonlinear model, 
\begin{equation}
G_l=
{\partial \hat{\bi{s}} \over \partial \lambda_l}
\bi{S}^{-1}\hat{\bi{s}}.
\label{glsl1}
\end{equation}
Here again the derivative has to be done on the actual reconstructed modes $\hat{\bi{s}}$. The interpretation of this derivative is as follows: we 
forward model the solution $\hat{\bi{s}}$ with a small change in 
forward model $\lambda_l$, $\bi{f}(\hat{\bi{s}},\lambda_{l,{\rm fid}}+ d\lambda_l)$, 
and then perform the optimization steps around the fiducial model 
to get the new solution $\hat{\bi{s}}_{\lambda_{l,{\rm fid}}+d\lambda_l}$. Taking the 
difference with respect to $\hat{\bi{s}}$ and dividing by $d\lambda_l$
gives us $\partial \hat{\bi{s}}/ \partial \lambda_l$. 
This can be numerically noisy, and analytic expressions for 
the derivative may give better performance. In the linear regime 
these are easy to derive, and perturbative methods can be used
to derive them beyond the linear theory. We will discuss an 
application to primordial non-gaussianity below. 

The remaining analysis follows the steps outlined in the previous section. 
We can combine the union of all
parameters into a vector $\bi{p}=(\bi{\Theta},\bi{\lambda})$.
Defining the fiducial values of all the parameters as $\bi{p}_{\rm fid}$ we 
can work with parameters relative 
to fiducial values $\Delta \bi{p}=\bi{p}-\bi{p}_{\rm fid}$.
The estimator for $\Delta \bi{p}$ parameters is then
\begin{equation}
(\bi{F}\Delta \bi{p})_l=G_l-b_l.
\end{equation}

Noise bias is given by
\begin{equation}
b_l=G_l(\bi{p}_{\rm fid},\hat{\bi{s}}_{s+n}),
\end{equation}
i.e. it is the estimator evaluated at the fiducial model simulation, 
guaranteeing the estimator is unbiased. 
The Fisher matrix can be computed as in the previous section 
by changing a single model parameter in the 
forward model and observing the response to all the parameters $\bi{p}$, both the
bandpowers and the model parameters.

\subsection{The general solution}

In general, a given parameter can affect both the primordial power 
spectrum and the forward model (e.g. matter density, massive neutrinos etc.). Similarly, 
one can work directly with the cosmological parameters instead of the
bandpowers. 
We can unify the methods of previous sections by treating all 
these cases with the same formalism.

Using $E_l$ in equation \ref{el} (where in general the sum is now over all modes $M$), 
with $\bi{\Pi}_l$ from equation \ref{pi} (replacing $\Theta$ with $p$),
and $G_l$ defined in equation \ref{gl} (replacing $\lambda$ with $p$), 
the total response is
defined as their sum $E_l+G_l$. 
We can now summarize 
the general analysis as following:

1) Use the
fiducial model to create a realization of the model $\bi{s}_{s}$
(or multiple realizations if needed), using methods described in previous section. 
Forward model $\bi{s}_{s}$ to the data domain to obtain $\bi{f}(\bi{s})$, and add noise to create a simulated realization of the data $\bi{d}_{s+n}$. Perform optimization to obtain reconstructed MAP $\hat{\bi{s}}_{s+n}$. Forward model 
again and compute derivatives with respect to forward model parameters
(equation \ref{lambda_deriv}). Define the noise bias as 
\begin{equation}
b_l=(E_l+G_l)(\hat{\bi{s}}_{s+n}).
\end{equation}

2) Perturb parameters $p_{l}$ by $\Delta p_{l}$, 
one at a time. The values of $\Delta p_{l}$ are to be small
enough so that they give a good estimate of the derivative, and large 
enough to reduce the roundoff errors. 
If $p_l$ changes the 
initial power spectrum $\bi{S}$ compute the derivative analytically to 
generate a new power spectrum
$\bi{S}^{\rm fid}+(\partial\bi{S}^{\rm fid}/\partial p_{l})\Delta p_{l}$. 
Generate modes
with mode amplitude
\begin{equation}
\vert \bi{s}_{s,l} \vert^2=\bi{S}^{\rm fid}+\left({\partial \bi{S}^{\rm fid} \over \partial p_{l}}\right)
\Delta p_{l}. 
\end{equation}
If $p_l$ changes the forward model include that in the 
forward model to generate the simulated data. Add the 
same noise as for the fiducial simulation. 
Perform the optimization to obtain MAP $\hat{\bi{s}}_{l,s+n}$, starting 
from $\hat{\bi{s}}_{s+n}$ for a fast convergence. 
Define the Fisher matrix elements as 
\begin{equation}
F_{ll'}={(E_l+G_l)(\hat{\bi{s}}_{l',s+n})-(E_l+G_l)(\hat{\bi{s}}_{s+n}) 
\over \Delta p_{l'}}. 
\end{equation}

3) Perform optimization on the data $\bi{d}$ using the fiducial model 
to obtain MAP $\hat{\bi{s}}$. Forward model 
using the fiducial model and compute the derivatives 
with respect to all the parameters in the forward model. 
The parameter estimators relative to fiducial model $\Delta \hat{\bi{p}}$ are
\begin{equation}
(\bi{F}\Delta \bi{\hat{p}})_l=(E_l+G_l)(\hat{\bi{s}}) 
-b_l.
\end{equation}
If the parameters are significantly different from zero (i.e. away from 
the fiducial model) repeat the procedure around the new fiducial model. 

The covariance matrix for $\Delta \hat{\bi{p}}$, 
both the 
cosmology and the model
parameters, is given by the inverse Fisher matrix. The solution 
can be multiplied by another choice of matrix $\bi{M}$. The estimator will be 
unbiased, but it may not be the minimum variance if some of the assumptions are 
violated. 
Additional evaluations are 
also necessary if one wishes to explore 
the posterior surface of $\Delta \hat{\bi{p}}$
beyond the multi-variate gaussian approximation 
given by the Fisher matrix. 

\subsection{Primordial non-gaussianity}

The procedure described above can also be applied to primordial 
non-gaussianity, which can be induced in the early universe, for 
example during the inflationary phase. In a typical scenario this 
creates a quadratic coupling between the modes, although higher 
order couplings are also possible. For example, 
the popular local non-gaussianity
model can be written as \cite{1990PhRvD..42.3936S}
\begin{equation}
\phi_{NG}(\bi{r})=\phi(\bi{r})+f_{nl}(\phi^2(\bi{r})-\langle \phi^2 \rangle ). 
\label{fnl}
\end{equation}
Here $\phi$ is the primordial gaussian 
potential, which can be evolved to late times 
using linear theory evolution, and related to linear density field 
through the Poisson equation and transfer function describing 
growth of structure through radiation and matter epochs. We can 
Fourier transform the equation above and convert to an observable density
field $\bi{s}$ (here meant to be linear density field, 
although we can also use other observables, such as 
the CMB temperature fluctuations) to obtain
\begin{equation}
s_{NG}(\bi{k})=s(\bi{k})+f_{nl}\left(\sum_{\bi{k}'} (T_{\Phi}s)(\bi{k}')
(T_{\Phi}s)(\bi{k}-\bi{k}')-\sigma^2\delta_{\bi{k}}\right),
\label{fnl1}
\end{equation}
where $T_{\Phi}(\bi{k})$ is the transfer function between the initial potential $\Phi(\bi{k})$ 
and linear density modes $s(\bi{k})$, $\delta_{\bi{k}}$ is the Kronecker 
delta being nonzero only for $\bi{k}=0$ mode, and $\sigma^2$ is the variance $\sigma^2=
\sum_{\bi{k}'} |(T_{\Phi}s)(\bi{k}')|^2$. We have written
the expressions as a sum over dimensionless modes, with the
appropriate volume factors inserted to convert from the usual
modes with dimension of square root of volume. 

One can view equation \ref{fnl1} as part of a forward model, where one 
first modifies the initial conditions using a deterministic expression 
of equation \ref{fnl}, 
with one free parameter $f_{nl}$, and then uses the forward model (such 
as an N-body simulation) to evolve the system forward into the data 
domain. One can therefore 
use equation \ref{glsl1}) as the estimator, 
with the derivative in equation \ref{lambda_deriv}
obtained by a numerical or analytic 
evaluation of the forward model at the MAP for 
two different values of $f_{nl}$. From equation \ref{fnl1} we have 
\begin{equation}
{\partial \hat{s}(\bi{k}) \over \partial f_{nl}}=
\sum_{\bi{k}'} (T_{\Phi}\hat{s})(\bi{k}')
(T_{\Phi}\hat{s})(\bi{k}-\bi{k}')-\sigma^2\delta_{\bi{k}},
\label{dfnl}
\end{equation}
which can be inserted into equation \ref{glsl1} (since we used dimensionless
version of modes $\bi{s}$, the power spectrum $\bi{S}$ should also be made dimensionless
by inserting appropriate factor of volume).
This gives us the optimal estimator for $f_{nl}$, and agrees in the 
limit of no noise and $T_{\Phi}=1$ with the estimator given in \cite{2007JCAP...03..019C} 
(expanding the estimator around $f_{nl}=0$), 
which was shown to be optimal in this limit. 
Note that the reconstructed modes $\hat{\bi{s}}$ are used in our 
estimator in equation \ref{dfnl}. This means the noise dominated modes will be suppressed, which 
is equivalent to inverse covariance matrix weighting of 
the data, guaranteeing the analysis is optimal in the presence of both the noise and the 
sampling variance. As argued in equation \ref{lambda_deriv}, one may be able to 
evaluate the derivative of equation \ref{dfnl} 
numerically, by running an $f_{nl}$ 
simulation using $\hat{\bi{s}}$. In the low noise limit this should 
work well, while its properties in the high noise regime need to 
be explored further. 

This method is an efficient
implementation of 
N-point function evaluation, 
which when done brute force involves
a summation of all possible 
products over N modes, where ${\rm N}=3$ in the case 
of example in equation \ref{fnl}. 
In our method we need 
a single
summation over all the modes (and a simulation to get the derivative). This is
still a 3-point function calculation, since the derivative 
with respect to $f_{nl}$ in equation \ref{glsl1} 
is quadratic in the field $\bi{s}$, and we multiply it 
with the field itself. 

One must also 
evaluate the Fisher matrix to obtain the errors and 
the correlation of
$f_{nl}$ estimator with all other parameters, as well as noise bias, 
following the steps in previous subsection. Specifically, we run a 
simulation with $f_{nl}=1$, and derive the response of the estimator 
in equation \ref{glsl1} as the difference between this estimator and the 
one for the fiducial model (which should be close to 0), 
which gives us the appropriate normalization  
$F_{f_{nl},f_{nl}}$. By the arguments given in section \ref{sec3} 
this also provides the inverse covariance matrix. 
Explicit evaluation of the likelihood function to quadratic order in $f_{nl}$
and evaluating the ensemble average of its second derivative
confirms this result \cite{2007JCAP...03..019C}. 
It is straight-forward
to evaluate the other parameter response to the difference between $f_{nl}=1$ and fiducial simulation. Among these parameters are the bandpowers, as 
well as the mean density parameter, which has to be determined from the 
data.
Our procedure thus gives the minimum variance estimator, together with the 
estimate of the corresponding Fisher matrix. 

The forward model parameters may not suffer from the 
sampling variance errors in the same way as the bandpower 
parameters. This depends on the 
full Fisher matrix and the mixing between the forward model parameters 
and the bandpower parameters.
The easiest way to see the sampling variance cancellation 
aspect is to have two different 
tracers, for example one that responds to $f_{nl}$ on large scales 
and one that does not, as in the 
case of local $f_{nl}$ at the linear level, where
these can be viewed as galaxies with bias $b>1$ and dark matter, respectively. In this case the dark matter will uniquely determine 
initial large scale modes $\bi{s}$, leaving the galaxies to determine $f_{nl}$ limited 
only by the noise of the former tracer. This is 
the basis of the sampling variance cancellation technique as applied to local 
$f_{nl}$ \cite{2009PhRvL.102b1302S}. Equations above should optimally 
extract all the information present in the data, 
automatically including all the information. 
At higher orders the sampling variance cancellation is implemented 
in the mode couplings: for example, a given large scale mode can 
have a variable coupling strength as a function of small scale modes in 
the presence of primordial non-gaussianity, allowing to identify the 
primordial non-gaussianity independent of the amplitude of the 
long wavelength mode. This is limited by the variance of the small 
scale modes. The method is thus not really sampling variance free, 
but limited by  the sampling variance of small scales modes, which
however may be small because
of the high number of modes.

\section{Application to dark matter simulations}
\label{sec5}

We have tested the bandpower method described in section \ref{sec3} on both 
2-LPT simulations and on N-body simulations. For the latter we use a cubic box of size $750\,\mathrm{Mpc}/h$ with a $128^3$ grid, using the publicly available code \textrm{FastPM} \cite{2016MNRAS.463.2273F} 
for simulating the nonlinear structure formation and the public code Cosmo++ 
\cite{Aslanyan:2013opa} for optimization.
We have used both the full N-body simulation and a simplified second order Lagrangian perturbation theory (2-LPT) for our truth, but below we only show results for the 2-LPT,
since the results 
were almost indistinguishable. 
We use standard Planck cosmology for all 
cosmological parameters and the outputs are evaluated at $z=0$. 

Generating a single realization of a full simulation 
using FastPM is fast. However, for the 
reconstruction one also needs a gradient with respect to all initial 
modes (equation \ref{rij}). This can be done using backward propagation 
\cite{2014ApJ...794...94W}, but is expensive, both because FastPM is slower than 2-LPT, 
and because gradient calculation is expensive and memory intensive. 
Moreover, as we show below, the optimization requires hundreds of calls 
to compute the gradient, making the full gradient considerably more 
expensive to compute. 

For the purpose of reconstruction in this paper we thus use a simpler method - we simply use 2-LPT multiplied with an appropriate transfer function. Specifically, we first compute a transfer function by simulating the density fields using PM and 2-LPT and taking the square root of the ratio of the power spectra. Then the gradients 
are performed using 2-LPT multiplied
by the transfer function. The 2-LPT gradients are given in appendix 
\ref{appc}. 
If the cross-correlation coefficient 
between full N-body and 2-LPT is close to unity then there is 
no loss of information in this process. In practice, at 
$k=0.3{\rm h/Mpc}$ the cross-correlation coefficient is 0.98
\cite{2012JCAP...12..011T}, and this approximation is extremely 
accurate over the range of $k$ we are interested in here. 

In this work 
we fix the transfer function, under the assumption that it does not vary 
with the input power spectrum. A full gradient implementation would 
remove this transfer function step.
The simulated fields are smoothed with a gaussian kernel of size $6\,\mpc/h$, which roughly corresponds to the pixel size. We add white noise to the truth simulation in real space. While smoothing was required to 
achieve convergence in our tests, with more accurate gradients 
one should be able to remove this step, 
and this is something we plan to pursue 
in the future. Smoothing can often be justified as a way to implement 
finite resolution of the experiment, although here it should be viewed 
simply as a tool to improve the convergence. 

Note that since we use a full N-body simulation for the 
truth, but we only use 2-LPT times the transfer function 
for the optimization, we cannot claim that 
the procedure extracts all the information optimally. For this reason we 
also performed the same analysis using 2-LPT as the truth: in this case 
no transfer function is needed, and within the context of 2-LPT 
dynamics this procedure is guaranteed to give the optimal power 
spectrum reconstruction. However, we observe almost no difference 
between the two cases, a consequence of smoothing, 
which eliminates signal relative to noise at 
high $k$, where the cross-correlation coefficient drops below unity. We only present 
results from the 2-LPT simulation as the truth. 

The forward model depends not only on the initial modes, but also on the 
assumed cosmology. Specifically, in the context of $\Lambda$CDM 
one needs to assume matter density $\Omega_m$, which determines the growth 
rate as a function of redshift. In this paper we simplify the analysis 
by ignoring this dependence, and we assume $\Omega_m$ is known. A 
more general analysis would explore the likelihood surface as a function 
of $\Omega_m$ as well, and we plan to explore this in the future. 

\begin{figure}
\centering
\includegraphics[width=\textwidth]{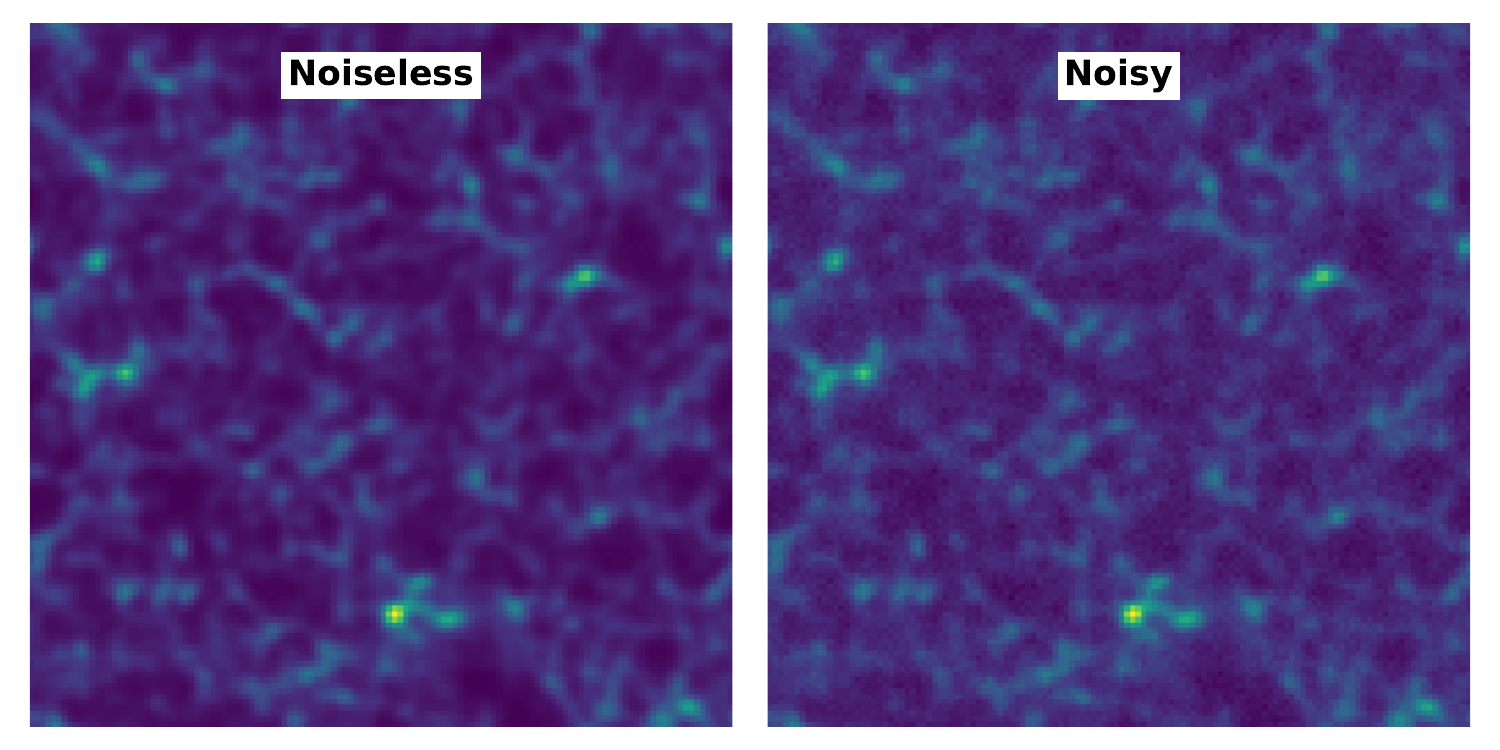}
\caption{The truth simulations without (left) and with (right) noise. The projection uses a slab with a thickness of 24 Mpc/h.}
\label{data_sim}
\end{figure}

\begin{figure}
\centering
\includegraphics[width=0.8\textwidth]{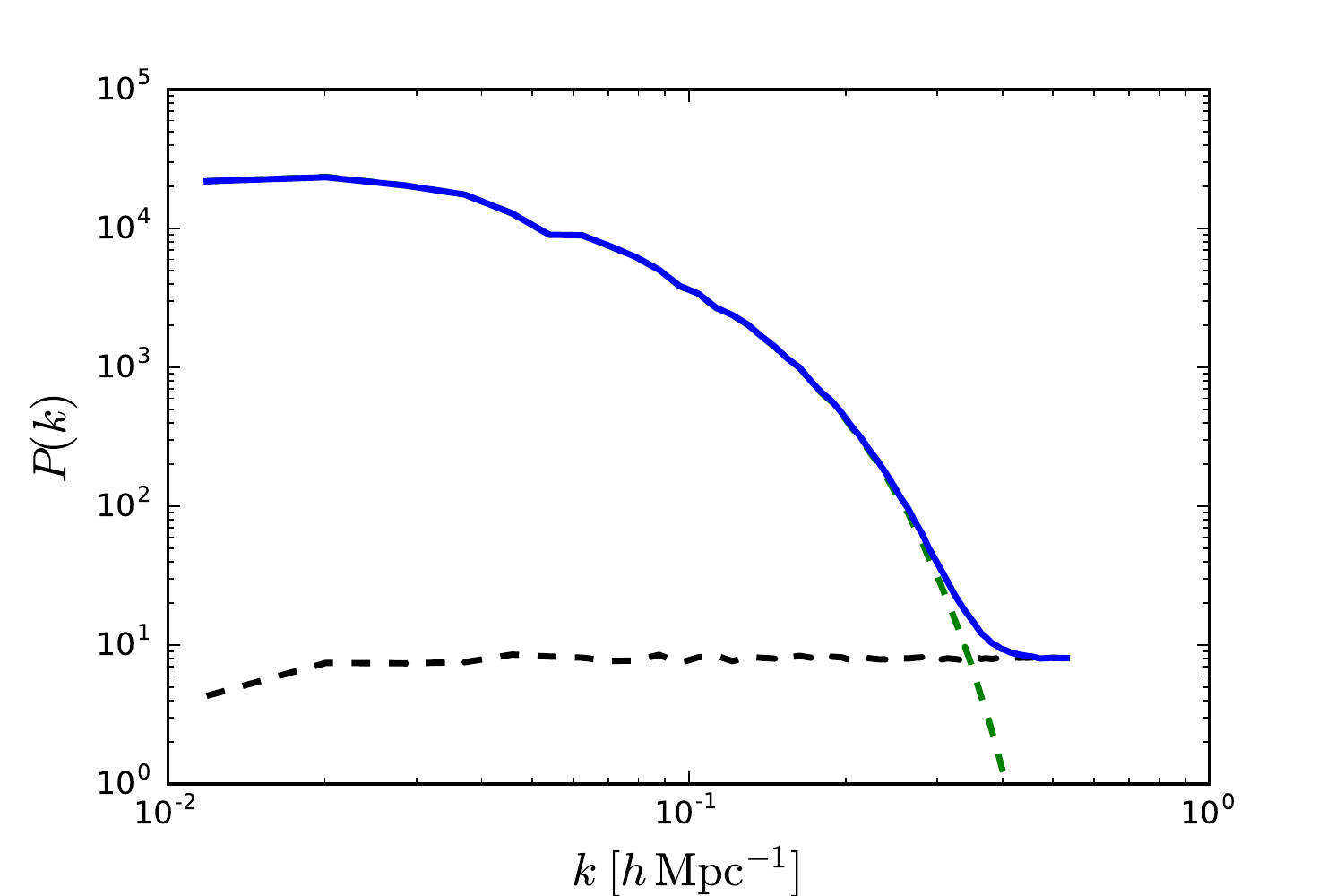}
\caption{The non-linear power spectrum of the truth simulation. The green dashed curve shows the simulation without noise, the black dashed curve is the noise power spectrum, and the blue solid curve is the total power spectrum.}
\label{pk_sim}
\end{figure}

We show a slice of our truth simulation, with and without noise, in Fig.~\ref{data_sim}. The nonlinear power spectrum of the truth simulation is shown in Fig.~\ref{pk_sim}. As a consequence of smoothing the 
data becomes noise dominated for $k\sim0.3\ihmpc$.

\begin{figure}
\centering
\includegraphics[width=\textwidth]{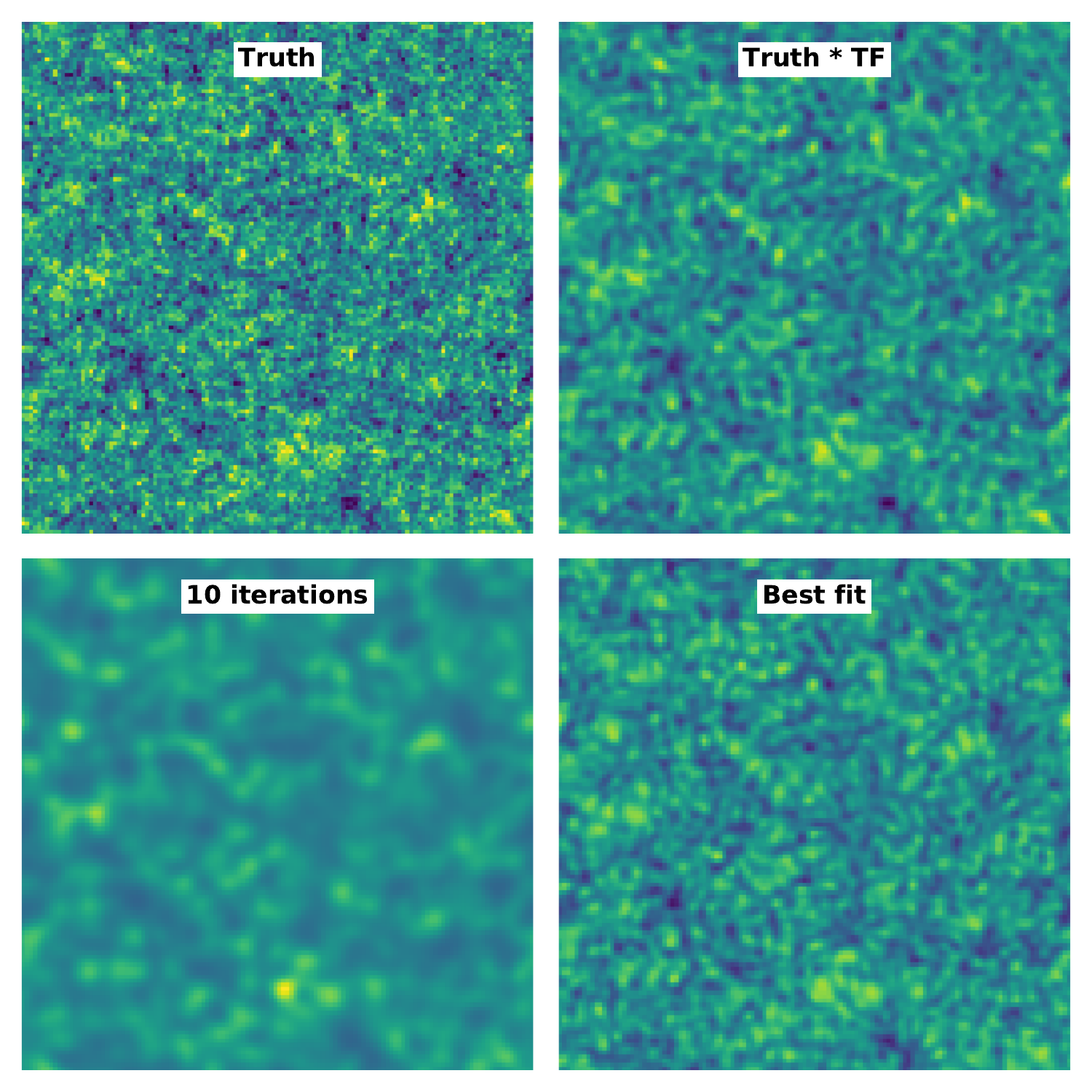}
\caption{The reconstructed initial linear density field (Best fit, lower right) compared to the truth simulation (Truth, upper left). Noise and smoothing prevent us having 
a perfect reconstruction, and small scale details are lost. We apply the transfer function on the truth simulation to demonstrate this (Truth*TF, upper right). Also shown are the reconstructed field at iteration $10$ (10 iterations, lower left). The projection use a slab with a thickness of 24 Mpc/h.}
\label{reconstructed}
\end{figure}

\begin{figure}
\centering
\includegraphics[width=\textwidth]{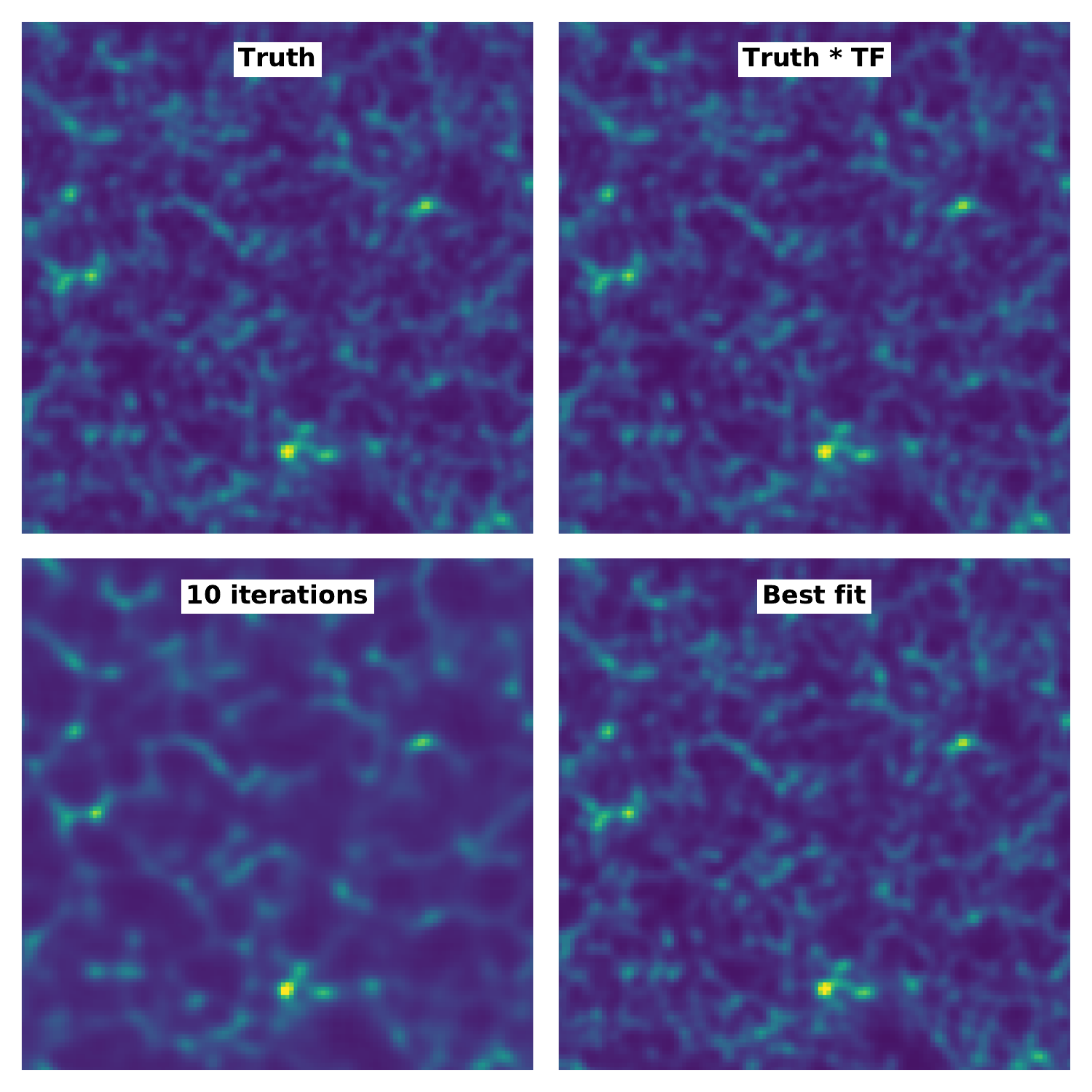}
\caption{Same as Fig.~\ref{reconstructed} but for nonlinear density field. Note that reconstructed and original fields appear much closer to each other. The projection use a slab with a thickness of 24 Mpc/h.}
\label{reconstructed_nonlin}
\end{figure}

\begin{figure}
\centering
\includegraphics[width=0.8\textwidth]{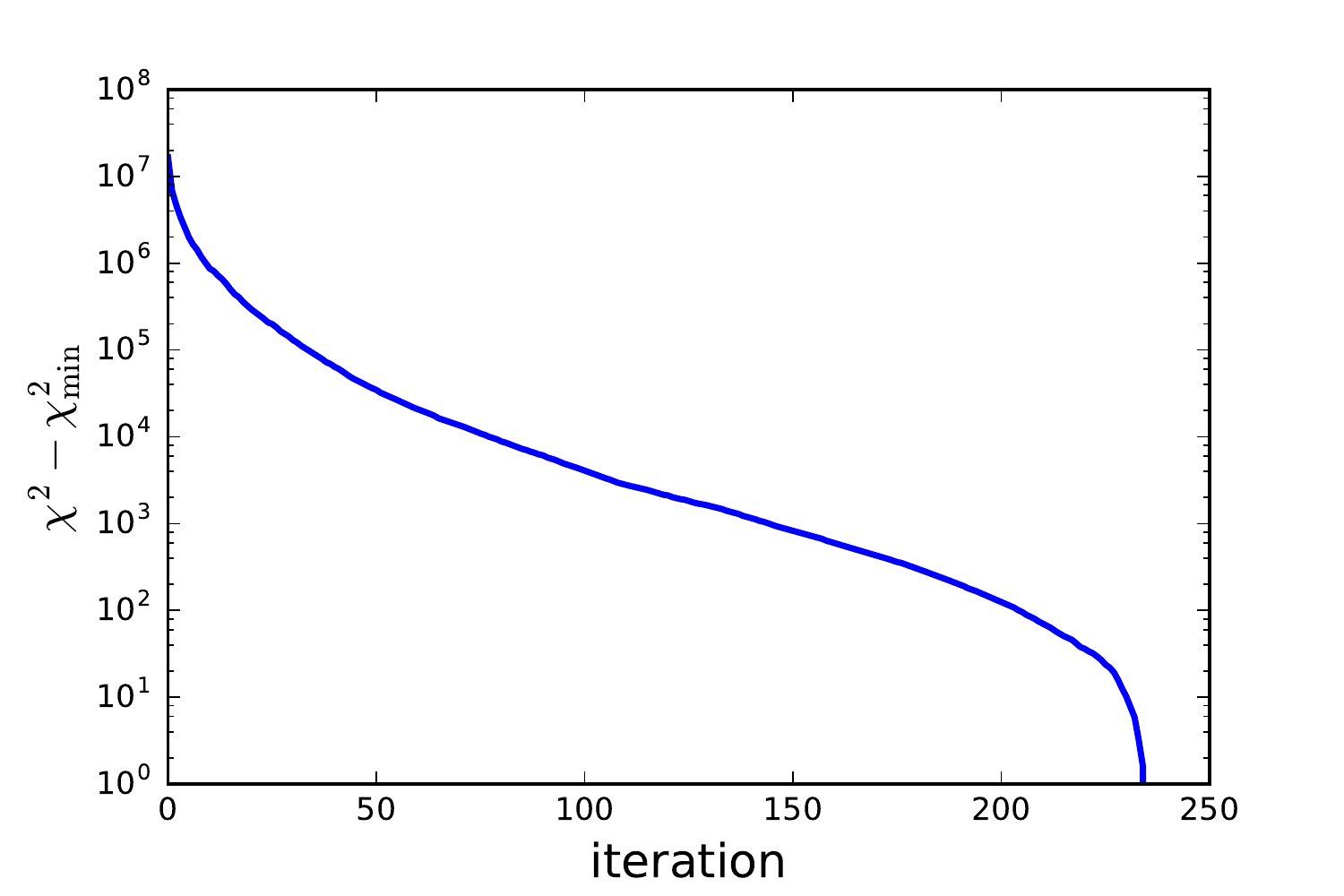}
\caption{$\chi^2-\chi^2_{\min}$ as a function of iterations.}
\label{chi2_fig}
\end{figure}

Let us first discuss the reconstruction of the initial density field. Our optimizer starts at the initial field of zero, and converges in $236$ iterations, and at each iteration we evaluate one or two line search steps, with a total of about 300 function evaluations. A slice through the original and the reconstructed initial density fields are shown in Fig.~\ref{reconstructed}. For comparison, we also show the reconstruction after $10$ and $100$ iterations. The corresponding nonlinear density fields are shown in Fig.~\ref{reconstructed_nonlin}. As we can see, the reconstruction works very well on large scales. However, the reconstruction is missing the small scale structure, which is expected, since the data is noise dominated on small scales and the prior, the first term in Eq.~\ref{chi2_eq}, drives the small scale modes to zero. The prior thus acts as a regularizer: in the absence of any real information in the MAP it sets the modes to zero. Lack of small scale modes is much more evident in the initial linear density field of Fig.~\ref{reconstructed}. However, it can also be noticed in the nonlinear field in Fig.~\ref{reconstructed_nonlin}. So far we have not encountered any problems with local minima: the procedure always converges to the same solution irrespective of the initial starting point. It remains to be seen whether 
that will remain the case when we have to deal with strongly nonlinear regime. 

Note that even after $10$ iterations the structures on large scales are already reconstructed fairly well. For this reason we would expect 
baryonic acoustic oscillations (BAO) to be reconstructed well after a few iterations 
only, similar to the BAO reconstruction methods \cite{2007ApJ...664..675E,2016arXiv161109638Z,2017arXiv170406634S}. 
After $100$ iterations the reconstructed field is indistinguishable by eye from the final iteration. 

To show the convergence rate of the optimizer we have plotted $\chi^2-\chi^2_{\min}$ as a function of number of iterations in Fig.~\ref{chi2_fig}. The 
convergence, here defined as a change of $\chi^2$ less than 1, is achieved
after 236 steps. After 150 steps the $\chi^2$ is about 1000 above the 
minimum. While this may sound a lot, the number of modes is $128^3 \sim 2\times 10^6$, 
so per degree of freedom this is a small change, and explains why visually 
the maps do not significantly change from 100 iterations to the 
final reconstructed map. We discuss this issue further below. 

\begin{figure}
\centering
\includegraphics[width=0.8\textwidth]{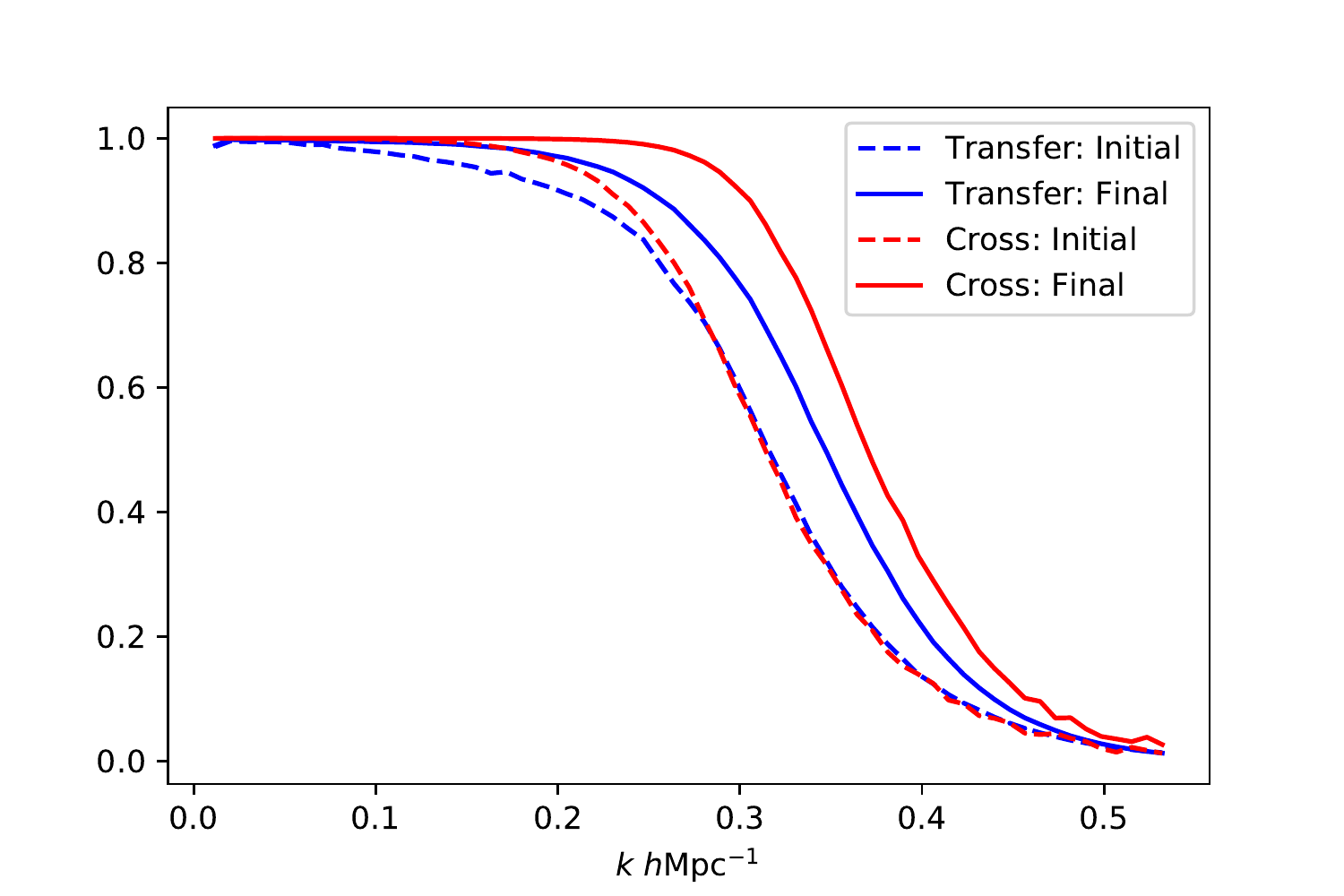}
\caption{The transfer function and the cross correlation coefficient between the reconstructed and the true density fields, as a function of $k$. The blue curves show the initial field, and the red curves show the final field. The dashed curves show the transfer function, and the solid curves show the cross-correlation coefficient. We see that the final density is reconstructed better than the initial density at the same $k$, a consequence of nonlinear power transfer from low $k$ to high $k$. 
}
\label{transfer_cross}
\end{figure}

To compare the reconstructed density field with the original truth field quantitatively, we plot the cross correlation and the transfer function between these two fields in Fig.~\ref{transfer_cross}. The transfer function is defined as the square root of the ratio of the power spectra, while the cross correlation is the ratio of the cross-power to the square root of the product of the individual power spectra. As expected, both the transfer function and the cross-correlation are very close to unity for large scale modes, and gradually drop down to $0$ for $k> 0.2\ihmpc$. This is expected, since the data becomes noise dominated beyond $k\sim0.3\ihmpc$. The effect of noise is 
more pronounced for initial density field. This is because high $k$ 
nonlinear density field is still significantly determined by low $k$
linear density, as a consequence of nonlinear mode coupling and 
power transfer. 

\begin{figure}
\centering
\includegraphics[width=0.8\textwidth]{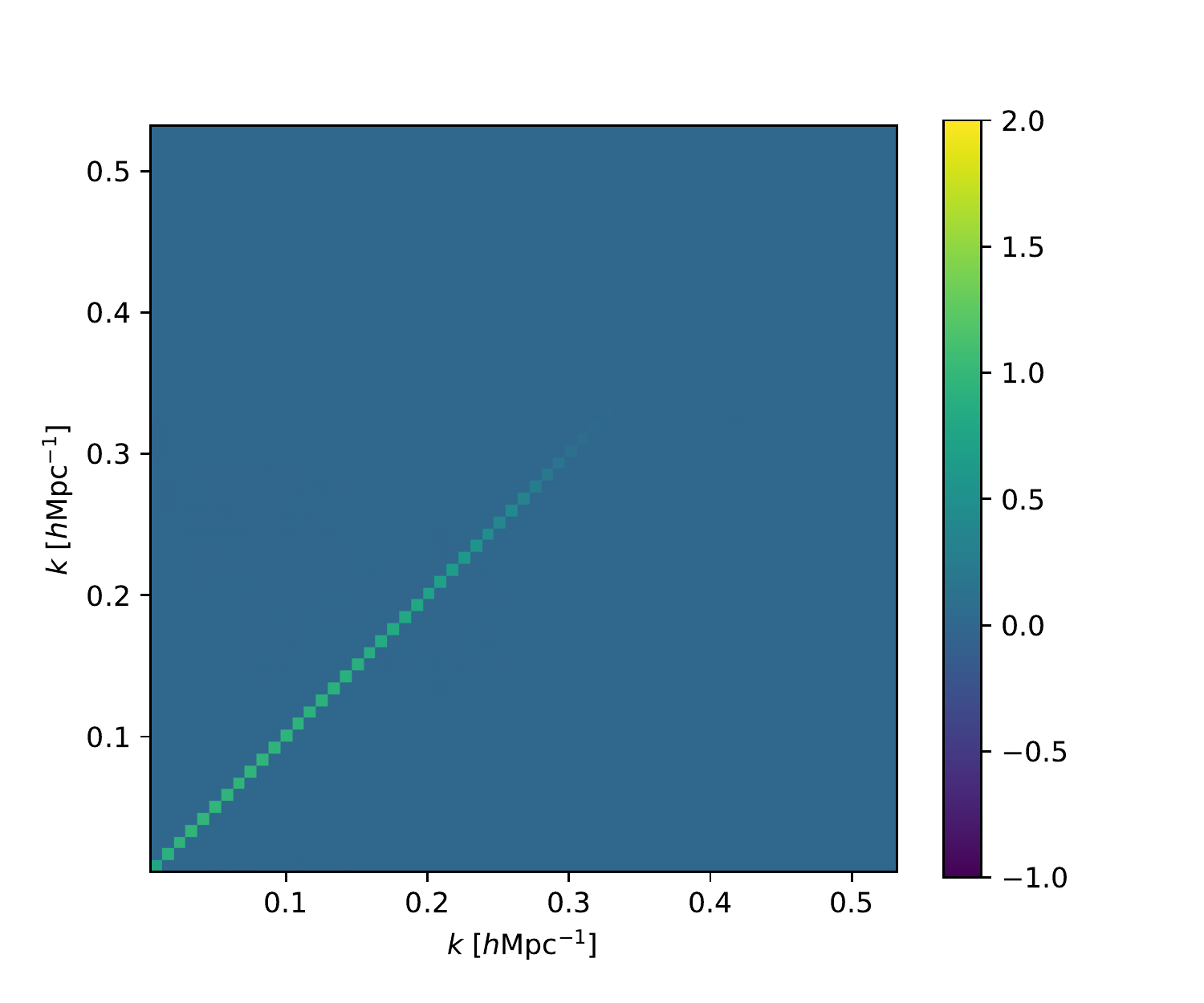}
\caption{The normalized Fisher matrix, defined as in equation \ref{t2}, 
computed from responses of the estimator to injection of power into a 
single bandpower bin. We observe no off-diagonal responses. At high k 
the responses vanish due to the noise.}
\label{window_slices}
\end{figure}

\begin{figure}
\centering
\includegraphics[width=0.8\textwidth]{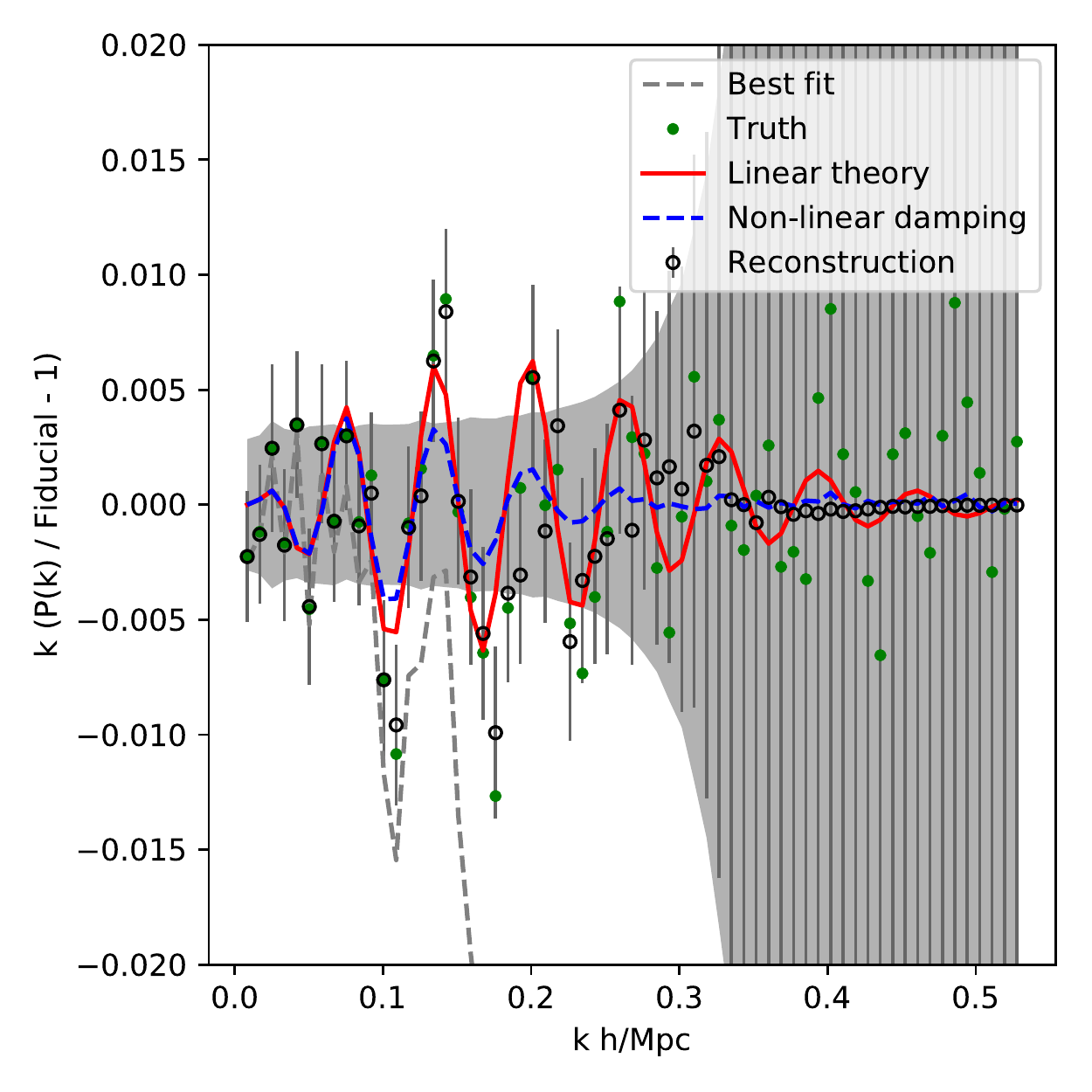}
\caption{The original and reconstructed linear power spectra, divided by no wiggle fiducial model. The red solid curve shows the original linear power spectrum which was used to simulate the truth. The gray dashed curve is the power spectrum at the best fit MAP mode reconstruction, which shows loss of power at high $k$ due to noise, and the black circles with error bars is our final power spectrum reconstruction corrected by the noise bias. The green points shows the actual realization of the linear power spectrum (simulation truth). The power spectrum is shown as the relative difference of all of the power spectra to the no wiggle power spectrum, enhanced by a factor of $k$ for better presentation. Also shown is the ratio of the nonlinear model power spectrum to the no wiggle one  (blue dashed curve). While nonlinear wiggles are completely damped for $k>0.2{\rm h/Mpc}$, linear ones are still visible, and
our reconstruction agrees with linear realization (simulation truth). The gray band shows size of the  error bars shifted to the center. The errors include sampling variance. The reconstruction appears perfect at high $k$ because we are using the same model for bias as 
for the truth, so for high $k$ where reconstructed MAP 
is zero this returns the fiducial model. 
In general the fiducial model would not be exactly the same as the truth, which is reflected by the size of the errors, becoming large at high $k$. }
\label{ps_reconstruction_fig}
\end{figure}

Let us now discuss the power spectrum reconstruction. We use $64$ equal width bins in $k$. For the reconstruction we use a fiducial power spectrum which is a no-wiggle version of the power spectrum for the truth simulation. This means that the BAO wiggles have been removed. One of the goals of the initial power spectrum reconstruction is to reconstruct the BAO wiggles without putting them into the fiducial model, 
and to show that the reconstruction completely removes nonlinear smoothing it is best if the fiducial power spectrum has no wiggles at all. 

We use equations \ref{fisher} and \ref{blmv} to compute the Fisher matrix
and noise bias. Specifically, we first create a fiducial power 
spectrum mock simulation as in equation \ref{dsim}, and perform exactly 
the same optimization steps as for the "truth" simulation. We next 
inject small amount of power into a single bandpower and measure the response
as given by all the other bandpowers. Here we start the optimization 
at the values given by the fiducial model simulation, and as a consequence 
we only need about 50 calls to converge. 
To make the procedure even faster we could do this for every N-th 
bandpower (e.g. N=5), since the mean transfer function squared of equation \ref{t2} 
is a smooth function of bandpower and can be interpolated across them. 
Once we have computed the Fisher matrix we also compute the noise 
bias using equation \ref{blmv}. 

We plot the normalized Fisher matrix (equation \ref{t2}) in figure~\ref{window_slices}. 
As we can see, the window functions are peaked
at the diagonal, and no significant mixing of modes occurs. This is expected when noise is 
low, since in that case we are reconstructing true modes. 
Note that we use periodic box, so there is no window function associated with the survey geometry itself: the nonlinear modes are orthogonal, so any window function deviation from a delta function would be
due to the nonlinear mode coupling to noise, creating off-diagonal elements. 
However, we observe no such mode coupling and no off-diagonal window function 
even for high noise. This result is intriguing and suggests that 
nonlinear evolution coupled to noise does not create much off-diagonal 
coupling. Our test case had a very low level of noise, and it 
remains to be seen if the same result is confirmed with higher levels of 
noise. Survey mask and other effects will create a window, which will 
be localized to the width of the bin of order $2\pi/R$, where $R$ is the 
size of the survey. It may thus
be possible to separate geometry effects 
from the nonlinear noise coupling effects, and develop rapid methods
for the estimation of the covariance matrix. Clearly this needs more detailed investigation, 
as it could greatly simplify the calculation of the window 
function and covariance matrix. 

Finally, we plot the original and the reconstructed power spectra in Fig.~\ref{ps_reconstruction_fig}, all relative to no wiggle fiducial power spectrum. 
We show the original truth power spectrum (red solid curve), the power spectrum of the reconstructed MAP initial density field (grey dashed curve), and our reconstructed power spectrum with error bars (black points). We can see that our reconstructed ``best fit'' MAP density field has a power spectrum that matches very well the ``truth'' power spectrum on large scales, but starts to decrease at small scales. As we discussed above, this is due to noise: in the presence of noise the optimal reconstruction sends the modes to zero. However, our reconstructed power spectrum (black solid curve) is 
able to determine the amount of lost power, and make the estimator 
unbiased. Indeed, it matches very well the model power spectrum on small scales, which is 
an artifact of the fact that we used the no wiggle version of the 
same power spectrum for bias term as for the 
truth. In reality we would not have this luxury so the results will be biased at high $k$, 
however not more than the computed errors, which properly account for this effect: we
see that the errors get very large for $k>0.35{\rm h/Mpc}$, since the transfer function 
of initial modes becomes very close to 0 and the MAP reconstruction is very close to zero. 

As we can see clearly from Fig.~\ref{ps_reconstruction_fig}, our reconstruction does reproduce the linear BAO wiggles, which match very well the wiggles in the truth power spectrum. For the nonlinear power spectrum the BAO wiggles are damped, as shown in the plot with the blue dashed curve. We can see from the plot that we are reconstructing the BAO wiggles of linear power spectrum, and not of the nonlinear power spectrum, achieving very good reconstruction. The last 
two wiggle with $k>0.35{\rm h/Mpc}$ are not reconstructed, because there the noise and 
smoothing already completely destroy the signal, and since we did not have BAO wiggles in 
our fiducial model they are not reproduced. The size of the errors reflects this, and the 
last BAO wiggle that is detectable is around $k>0.3{\rm h/Mpc}$. 

As discussed in section \ref{sec4} it is not guaranteed that our procedure gives a reliable Fisher matrix that can be used as the 
covariance matrix, although this is guaranteed in the linear regime and in the low noise regime. We have computed goodness of fit value of the estimators relative to the truth 
$\Delta\bi{\Theta}^{\rm truth}$ (which is not 0 because our assumed fiducial 
model is not the truth model)
\begin{equation}
\left[\bi{W}\Delta\bi{\Theta}^{\rm truth}
-\Delta\hat{\bi{\Theta}}^,\right]^{\dag} \bi{C}^{-1}\left[\bi{W}\Delta\bi{\Theta}^{\rm truth}-\Delta\hat{\bi{\Theta}}^,\right] \sim 38,
\end{equation}
using 40 bandpowers up to $k_{\rm max}=0.4{\rm h/Mpc}$. The reduced $\chi^2$ of 0.94 suggests the covariance matrix is reliable. 

\begin{figure}
\centering
\includegraphics[width=0.8\textwidth]{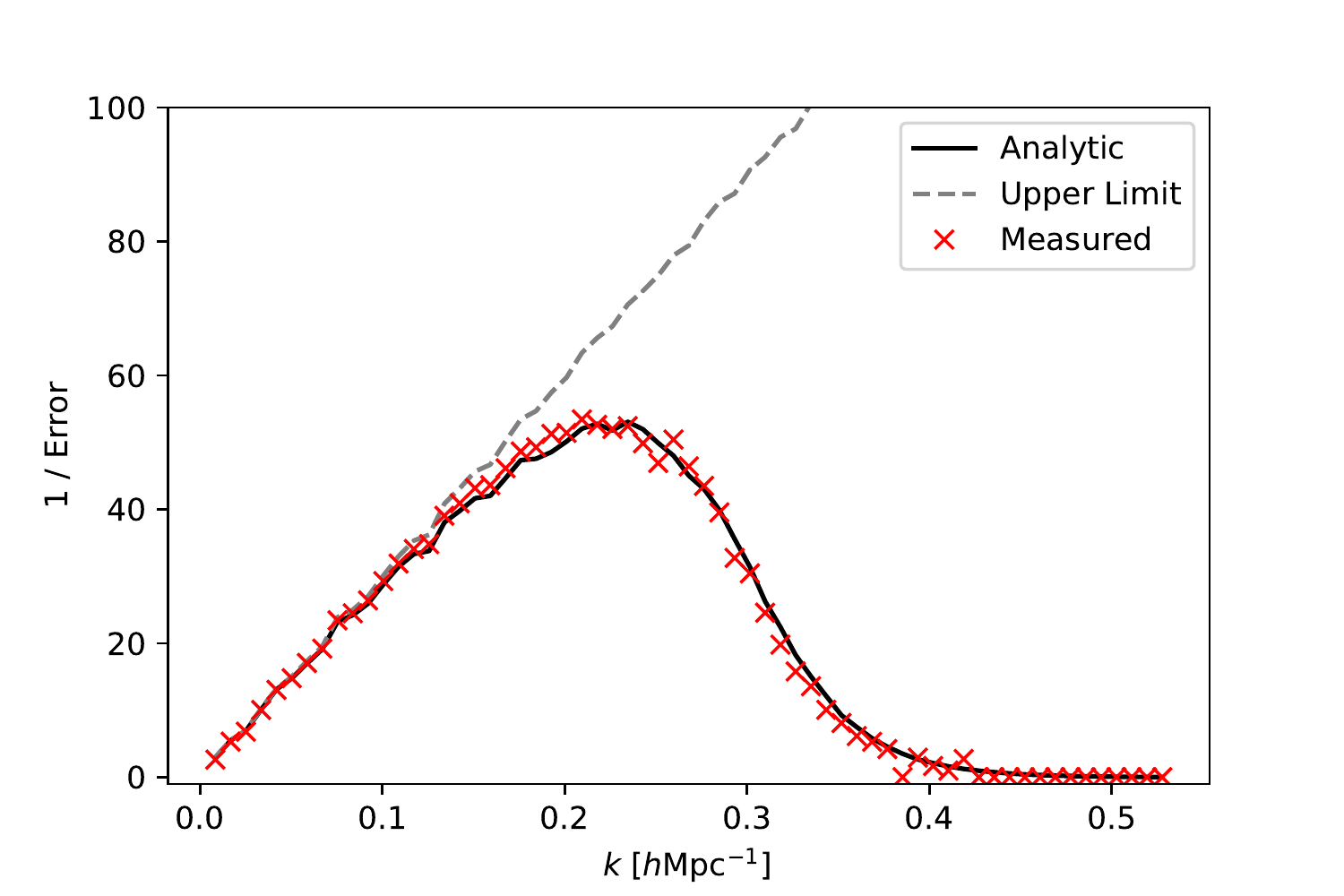}
\caption{The square root of the diagonal terms of the Fisher matrix for 
normalized parameters, 
which is the signal to noise for the bandpower. We compare our main 
method (red crosses) to analytic prediction from equation \ref{fllel2} (solid black line), finding a good agreement between the two. We also plot the maximal inverse error in the case of perfect reconstruction (dashed line).} 
\label{fisher_diag}
\end{figure}

We have also 
compared the diagonal terms of the Fisher matrix to the 
simple approximation of equation \ref{fllel2}, 
finding good agreement between the two methods (figure \ref{fisher_diag}). 
This, and the nearly diagonal window (figure \ref{window_slices}) suggests 
that one may be able to estimate a reliable covariance matrix even 
without doing any simulations (more precisely, with a single simulation to 
compute $\bi{b}_l$, making 
sure the estimator is unbiased). This clearly deserves a more detailed 
investigation, in particular 
in the presence of a survey mask and other complications, 
but is beyond the goal of this paper. 

\subsection{Convergence and computational cost}

Let us look at the computational costs of our method. 
As mentioned above the cost of a single optimization procedure is 
about 230 forward model calls, each being a few FFTs (to compute 2-LPT) 
of the $128^3$ mesh. On the data this is done once. For the simulations 
we have to do the same for the fiducial simulation realizations, 
and then for each of the Fisher matrix rows, where we inject a small 
amount of power into a single bandpower on top of the fiducial power. 
The latter would thus appear to dominate the computational cost, but 
because the Fisher matrix can be interpolated across 
the bandpowers (we only evaluate every fifth row), 
and because we can start the optimization from the 
fiducial model solution, which reduces significantly the number of iterations, 
in practice this entire operation is comparable in cost to 
the optimization of fiducial model simulation, or of the data. 
To achieve required precision one also needs to repeat the procedure 
until required precision is achieved. In our tests we have not found the 
need to average over many realizations. 

Another possible acceleration is to extract the results 
without reaching the global minimum. As figure \ref{reconstructed} shows
the solutions between 100 iterations and final convergence are almost indistinguishable, since 
the optimizer is changing a large number of modes by a small amount that cannot change the
solution significantly. This result is confirmed by the actual value of $E_l$ as a 
function of iteration number, which hardly changes at higher iterations even when the loss function is still changing 
significantly. 
In general, if we have $M$ modes $\bi{s}$ and the optimizer has determined each
to within $\epsilon \sigma$, where $\sigma$ is the error for individual mode, then for $\epsilon \ll 1$ 
there is no loss relative to perfect optimization solution, in the sense that the total error is 
given by $(1+\epsilon^2)^{1/2}\sigma$. As long as the error is stochastic the same will be true 
for the bandpower error, where we average several modes. But this additional noise adds $\epsilon^2M$ to the 
overall loss function, and for large $M$ this can be a large number even if $\epsilon$ is small. 
For example $\epsilon=0.1$ only increases the overall error by 0.5\%, suggesting that even 
the optimization solutions which are away from the global minimum by a large number may still be good enough. Another way to understand this argument is 
through the sampling perspective: when sampling each mode is on 
average $1\sigma$ away from MAP, and most of the sampling is done in a 
narrow range of loss function being $M \pm \sqrt{M}$ above the global minimum. 
The sampler never descends down to the minimum, and its exact position is 
irrelevant for the posterior distribution of the modes, which is ultimately 
what determines the solution. 
Figure \ref{chi2_fig} suggests that we could stop the optimization after 50-100 steps and still satisfy the 
condition that $\epsilon^2 \ll 1$. In practice we observe strong 
correlations in the modes of a single bandpower 
along the optimization path, so the argument above is too naive
and one must develop a more robust criterion when to stop. We also 
observe this argument to fail when computing Fisher matrix elements, 
where small differences between solutions 
are relatively more important. 

A related issue is the question of convergence to the global minimum. The 
loss function is only convex for the linear problem. For the nonlinear 
problem it becomes non-convex, meaning that there can be multiple 
local minima. In the mildly nonlinear regime this will not happen, 
but in deeply nonlinear regime this is very likely to happen. Changing 
the starting point of optimization, or using a stochastic descent which 
tries different paths down the hill are typical 
approaches to this problem, but a general solution is difficult. 
However, as argued above a local minimum is only problematic if it is 
very far from the global minimum. A situation where there are many 
high quality local minima at the bottom of the loss function is not of 
much concern, since all those solutions are in some sense equivalent, specially 
if they all give the same transfer function. 

Another numerical issue are saddle points. 
These are numerically problematic, since 
they are attractors for Newton's method optimizers: they 
may give an impression that  
convergence has been reached, only 
for the optimizer to suddenly 
start reducing its loss function again 
once it has evolved past the saddle point. 
We have observed this in the configurations with high noise. 
Adding noise or momentum to the gradient, taking absolute value of the Hessian eigenvalues
or some 
other way to change the Hessian make these issues less problematic \cite{DBLP:journals/corr/DauphinPGCGB14}. 
More work is needed to develop optimal optimization methods for our problem. 

In any case, the large number of forward model evaluations, in the hundreds to thousands, and the need to compute the gradients
puts a high demand on its computational 
cost, and approximate 
methods like 2-LPT or fast simulations like FastPM \cite{2016MNRAS.463.2273F}
are crucial for the success of this program. 
We note that FastPM with 10 time steps is 7 times slower
than 2-LPT, while its transfer function with full N- body simulations at $k=0.3{\rm h/Mpc}$ is 
1.02, relative to 1.2 for 2-LPT, so its computational cost is not prohibitive, while dramatically improving 
the accuracy. However, one also needs to demonstrate 
that its gradients can be implemented accurately and can be evaluated. In practice the choice of 
forward model will depend on the level of noise and resolution in the data, with low resolution/high noise data requiring only 2-LPT and high resolution/low noise data requiring N-body simulations with low number of time steps. 
It remains to be seen whether this will 
change with a more realistic survey geometry, where window functions will be less compact. 
It may be possible to 
reduce this number further if the convergence of the optimizer is accelerated.
We note that in our tests a simple conjugate gradient optimizer was a factor of 2-3 times 
slower than L-BFGS, which is the reason we only present the latter in these tests. We plan to investigate some of these acceleration issues further in the future. 

\section{Conclusions}
\label{sec6}

The purpose of this paper is to develop a formalism 
to optimally extract cosmological information in the nonlinear regime, and 
to develop numerical methods that are computationally feasible and 
preserve the near optimality of the analysis,
and allow an implementation of this formalism both in the linear and nonlinear regime. 
Traditional CMB/LSS analyses start with the power spectrum or correlation 
function of the observed field (such as CMB temperature and polarization, 
weak lensing shear or galaxy positions), 
and often end there. Even in this case the two-point function analysis is typically not optimal, because the data are not inverse covariance 
matrix weighted.

In the nonlinear regime the LSS evolution creates non-gaussian signatures. 
This is a generic term 
that contains a lot of different manifestations: they range from the
higher order correlations such as bispectrum, to existence of
special objects like the density peaks (e.g. clusters), 
to non-trivial topologies such as sheets, filaments and voids. 
The problem with these descriptions is that they are not 
exhaustive, i.e. it is not possible to prove that any specific
combination extracts all the information. Indeed, often it is not 
even clear that these additional statistics increase 
the amount of information relative to the power spectrum 
at all. When combining different statistics one must also obtain their joint covariance matrix, which is a challenge in itself, since it typically 
requires a large number of mock data realizations (parametric bootstraps). 

In this paper we present an (approximately) Bayesian analysis 
that attempts to nearly optimally extract 
the information contained in the data given a 
nonlinear structure formation models of LSS. By writing down an 
exact probabilistic model the optimality of the analysis is guaranteed 
as long as all the steps are solved exactly. 
In our approach we do not attempt to produce 
an exact solution, which is prohibitively 
expensive, but instead we develop 
a series of controlled approximations that we argue approximately 
preserve the optimality of the solution.
Should these approximations turn out to be insufficient one can 
attempt to generalize the approach used here. 

Specifically, our approach 
is to expand the likelihood around the maximum a posterior (MAP) 
solution for the initial modes, and integrate out the modes around this 
solution. 
In our specific implementation we assume existence of a single global MAP
solution. 
This assumption is guaranteed in the linear model, and 
we argue that it is likely to be the valid at least in the mildly nonlinear 
regime, which is our primary focus. We do not analytically compute the 
integration volumes, which correspond to the product of all variances of 
the modes, but instead rely on simulations to compute their 
derivative with respect to the parameters. We 
do this using simulations, and the realization dependence of the 
variance is lost. We 
have argued that at the linear level this is a valid and lossless approach. 
We also expect any corrections to be small in the low noise limit, and 
in general we expect this to be a very good approximation. 
In this context it should be stated 
that in the presence of multiple local minima the solutions should not be expanded around the global MAP, but around the one 
that maximizes the loss function times the log of posterior volume (equation \ref{marg}), which differs from what MAP maximizes by 
$\det \tilde{\bi{D}}^{-1/2}$. While we have argued that it is likely 
that one does not need to average over multiple local maxima, this 
does not imply that finding the one that maximizes MAP (or volume 
corrected MAP) is easy. 
A proliferation of lower quality (higher loss function) saddle points 
can still make convergence slow, and there is no guarantee that the 
optimization will find a global minimum, as opposed to a local minimum 
far from the best solution. We have argued that if these local maxima are 
close to the global MAP and all have the same transfer function 
it does not affect the unbiasedness of the solution. 

Similarly, the optimality of the method is only guaranteed 
if the forward model is exact, even if it is unknown (exact in the sense that 
the nuisance parameters always cover the truth). 
Formally one adds more and more forward model parameters, and marginalizes
over them. 
While it is easy to achieve this on 
large scales, it becomes increasingly more difficult to do so on 
small scales where many uncertain physical processes can affect the data, 
and one needs to include all possible variations consistent with the 
allowed physical processes, including physical priors on these parameters. 

We argue that the bandpowers of optimally reconstructed initial 
power spectrum are lossless summary statistics, at least for the 
simplified case where the forward model connecting initial modes
to the data is unique. 
We provide a derivation of the bandpower estimators, the corresponding Fisher matrix and noise 
bias, which fully quantify the information content of LSS data. 
Note that our approach does not 
involve any sampling of the modes, and we believe that current 
implementations of sampling procedure to be too 
slow to be of practical use in these applications. 

We then generalize the analysis allowing also for parameters that control 
the mapping from the initial modes to the final data, both 
cosmology parameters such as matter density or massive neutrinos, 
and astrophysical modeling 
parameters such as galaxy bias, shear bias etc. We expect that these 
parameters in general are not degenerate with the power spectrum 
amplitude parameters, even if they may be at the linear level: 
in the nonlinear regime this degeneracy can be broken, since nonlinear 
gravitational effects cannot be exactly mimicked by the nuisance parameters. 
This allows for possibility of marginalizing 
over the nuisance parameters internally, without the need for external 
constraints. 
We show that one can treat primordial non-gaussianity
as a forward model parameter and present an optimal analysis method 
to extract it. Our method
involves only a single sum 
over the data, multiplied by a response function numerically derived 
at each data point. 

For linear models, such as the
CMB or the LSS on large scales, where the relation between the data and the modes is a simple Fourier transform, the
MAP is guaranteed to be at the global 
maximum with a gaussian posterior and this guarantees optimality \cite{1998ApJ...503..492S}. Even in this case 
we expect the numerical methods developed here, optimization 
and fast Fisher and noise matrix evaluations, to be useful 
in applying these estimators to the real data. We will present 
our results of applying these methods to weak lensing and CMB applications elsewhere. 

This paper focuses on the theoretical 
formalism, deriving the summary statistics 
in the context of nonlinear structure formation models.  
We implement the procedure on a simplified, yet nontrivial test case, 
where we use 2-LPT or N-body simulations as a forward model to generate a 
nonlinear density field, while we use
simplified 2-LPT times the transfer function 
for gradient evaluations during the optimization. 
We show that the resulting solution recovers the linear baryonic acoustic 
oscillations, achieving a near optimal BAO 
reconstruction. Since our gradients are not exact we do not 
claim that the reconstruction is optimal, although this is very
likely to be the case for BAO (since 2-LPT is expected to fully model 
nonlinear BAO evolution). 
Our reconstructed power spectrum is unbiased and comes with the bandpower 
window matrix and the (disconnected) covariance matrix that describes well 
the solution. 

Some of the more realistic effects that remain to be included are incomplete data coverage, nuisance parameters such as the biasing 
of galaxies, and baryonic effects on the matter distribution. 
Equally importantly, while for the weak lensing the noise model is simple, for galaxy data noise probability distribution needs 
to be understood better, and sub Poisson noise should be achievable \cite{2009PhRvL.103i1303S}. When applying to weak lensing one additional 
complication is the elongated radial projection window, which does not allow for a full 3-d reconstruction even when tomographic reconstruction 
is used, and will lead to significant degeneracies 
in the radial direction. 

Ultimately we expect the methods developed here to
superseed the traditional data analysis methods in cosmology. When this will 
be achieved depends mostly on the remaining technical and computational challenges that need to be addressed. 
Among these are fast evaluations of
derivatives of N-body simulations, where we expect FastPM code \cite{2016MNRAS.463.2273F}, which achieves percent level agreement of halo clustering against 
high resolution simulations with as few as five time steps, 
to be particularly suitable for this purpose. Another challenge is 
the ability to perform gradients of galaxy density field, as well as
gradients of forward model parameters evaluated on MAP. We plan to 
investigate some of these issues in the future. 
Given the higher computational cost of this 
approach, and various numerical challenges in the nonlinear regime,
there may still be a long road ahead before we achieve this goal, but 
the payoff will be an increased precision of cosmological constraints
and (near) optimality of the analysis.

\acknowledgments

We thank Emanuele Castorina, Simone Ferraro, Ryan Giordano, Patrick McDonald, 
Martin White and Matias Zaldarriaga for useful discussions. We acknowledge support of NASA grant NNX15AL17G. This research used resources of the National Energy Research Scientific Computing Center, a DOE Office of Science User Facility supported by the Office of Science of the U.S. Department of Energy under Contract No. DE-AC02-05CH11231.

\appendix

\section{Appendix: Beyond the gaussian posterior approximation}
\label{appa}

The power spectrum bandpower procedure described in the main text is explicitly unbiased, 
and provides bandpowers and their covariance matrix under the gaussian approximation for the 
posterior. 
In a typical application to LSS the number of modes in a bandpower is large, 
so that by the central limit theorem the gaussian approximation to the likelihood in equation 
\ref{llik}
suffices. When the number of modes inside the bandpower is small this is no longer the case: 
a model with a large amount of power relative to the best estimate also has a larger 
sampling variance error, making this model more likely 
than the gaussian approximation would suggest. Since this situation is most likely to 
occur on large scales, where the modes are linear, we will consider linear case
only. For the simplest linear model with complete coverage one can write 
the likelihood exactly in 
terms of $\hat{\bf\Theta}$ (${\bf\Theta}$) as 
(inverse) Wishart distribution \cite{2009PhRvD..79h3012H}. For our purpose we need $L({\bf\Theta}| \hat{\bf\Theta})$,
\begin{equation}
-\ln L({\bf\Theta}| \hat{\bf\Theta})= \sum_l X_l\left( x_l-\ln x_l -1\right),
\label{invw}
\end{equation}
where 
\begin{equation}
x_l= {\hat{\Theta}_l+b_l \over \Theta_l+b_l}, 
\end{equation}
and $X_l=K_l/2$, where $K_l$ is the number of modes in the bin. For convenience 
we have normalized the log-likelihood to zero when $x_l=1$. 

A simple generalization for the optimal quadratic estimator 
is to use equation \ref{invw} together with
the uncorrelated version of the estimator in equation \ref{mdec}. 
Since the product of inverse square root of Fisher matrix with the 
Fisher matrix is the square root of the Fisher matrix, this requires
its evaluation. 
The best option is for matrix ${\bi F}^{-1/2}$ to be symmetric, which 
can be evaluated by diagonalizing ${\bi F}={\bi{ZDZ}^{\dag}}$, and then evaluating the square root of the diagonal elements to give ${\bi F}^{-1/2}=\bi{Z D}^{-1/2}\bi{Z}^{\dag}$. This will make the window function narrower (for narrow windows, by Taylor expansion, 
the off-diagonal terms will be approximately halved). 
Once diagonalized, we can use 
\begin{equation}
X_l={ \left((\bi{MF\Theta})_l+b_l\right)^2 \over (\bi{MFM})_{ll}}.
\end{equation}
One can interpret this term as giving an effective number of modes measured
in bandpower $l$ (divided by 2), given the noisy and incomplete data. The effective number of modes
is not only a function of volume coverage, but also of noise and fiducial power: 
pixels that are noisy relative to the fiducial power are downweighted, reducing 
the effective volume. Note also that the signal to noise in bandpower $l$ 
is lower than $X_l^{1/2}$, because of noise bias $b_l$,
\begin{equation}
\left({S \over N}\right)_l= {(\bi{MF\Theta})_l \over (\bi{MFM})_{ll}^{1/2}}.
\end{equation}

\section{Appendix: Alternative approaches to 
Fisher matrix and noise bias}

\label{appg}

Fisher matrix $F_{ll'}$ is defined as the response of the modes in a bandpower  bin $l$ to modes in a bandpower bin $l'$. 
To evaluate this response we 
can compute the cross-correlation of one mode of bin $l$ to the modes in bin $l'$. We label this mode with 
index $k_l$, and there are $K_l$ modes inside the bin $l$. 
We define a transfer function between initial and reconstructed modes  as a cross-correlation divided by auto-correlation
\begin{equation}
T(k_l,k_{l'})={\left\langle \hat{s}_{s+n}(k_{l})s_{s}^{*}(k_{l'}) \right\rangle \over 
\left\langle\vert {s}_s(k_{l'}) \vert^2 \right\rangle}=
{\left\langle \hat{s}_{s+n}(k_{l})s_{s}^{*}(k_{l'}) \right\rangle \over S_{k_l'}}.
\label{tkk}
\end{equation}
When maximally correlated and without noise this equals unity for $k_l=k_{l'}$. 
By defining this as a 
transfer function and dividing with the actual power we reduce the 
sampling variance. 
We can also choose the actual power for each mode $s_s$ to 
be fixed to the fiducial power, reducing fluctuations further. 
This is different from the typical mock (Monte Carlo) simulations, which require more realizations to converge. For the same reason our noise fluctuations are
forced to have the same variance via random $Z_2$ realizations. If 
necessary 
the procedure is repeated on several realizations, until we 
achieve the required convergence. 
 
Fisher matrix is then given by 
\begin{equation}
F_{ll'}=\sum_{k_l,k_{l'}} {\vert T(k_l,k_{l'}) \vert^2 \Pi_l \Pi_{l'}
\over 2S_{k_l}^2} .
\label{fisher_transfer}
\end{equation}

To evaluate the noise bias we use the difference between the auto-correlation, which contains noise contribution,
and cross-correlation, which does not,
\begin{equation}
b_l= {1 \over 2}\sum_{k_l} (S_{k_l}\Theta_l)^{-1}\left\langle\vert \hat{s}_{s+n}(k_l) \vert^2 \right\rangle -\sum_{l'} F_{ll'} \Theta^s_{l'},
\end{equation}
where
\begin{equation}
\Theta^s_{l'}=K_{l'}^{-1}\sum_{k_{l'}} \left\langle \vert {s}_s(k_{l'}) \vert ^2\right\rangle .
\end{equation}

From this we finally obtain the unbiased 
estimator by adding over all $K_l$ modes within the bandpower $l$
\begin{equation}
\hat{\Theta}_{l}={\sum_{k_l}(S_{k_l}\Theta_l)^{-1}\vert \hat{s}(k_l)\vert ^2-2b_l \over 2\sum_{l'} F_{ll'}}.
\label{hattheta}
\end{equation} 

This is our minimum variance estimator in the sense of equation \ref{hatth}, i.e. with $M_{ll'}=\delta_{ll'}(\sum_k F_{lk})^{-1}$ definition. 
It is manifestly unbiased, in the sense that $\langle \hat{\Theta}_l \rangle=\sum_{l'}W_{ll'} \Theta_{l'}$ (equation \ref{covfish}). 
It is quadratic in the optimal field
reconstruction $\bf{\hat{s}}$, which is in 
turn nonlinearly related to the data $\bf{d}$. 
The answer is given by the optimal field reconstruction, corrected for the noise bias and properly normalized. 

The procedure  is useful to clarify the nature of noise bias 
and Fisher matrix, but its convergence is very slow. The reason is 
that we have to compute cross-correlations between all the modes, 
then square it, and this process will be biased positive even if there 
are no correlations between the modes, and will only 
slowly converge to the correct value with the number of simulations. 
Moreover, the calculation of the Fisher matrix using \ref{fisher_transfer} is quadratic in the total number of modes, which can become prohibitive for $~10^6$ or more modes. In the main text we developed 
an alternative approach that converges faster. 

A second alternative method takes advantage of the 
Fisher matrix interpretation as the 
covariance matrix. Suppose that the Fisher matrix is diagonal. This 
is true for uniform noise and periodic box simulations used in this 
paper (figure \ref{window_slices}). In this case we have from 
equation \ref{wfmv}
\begin{equation}
F_{ll}^2\rm{Var}(\Delta \Theta_l)=\rm{Var}(E_l), 
\end{equation}
where $\rm{Var}$ is the variance of the quantity, ie $\rm{Var}(x)=\langle x^2\rangle -\langle x \rangle^2$. 
Assuming $\hat{s}_{k_l}$ are gaussian distributed we can use Wick's theorem 
$\rm{Var}(E_l)=\rm{Var}(\sum_{k_l} \hat{s}_{k_l}^2)=2E_l^2/K_l$. This, 
together with equation ref{minvar} gives equation \ref{fllel2}.
As shown in figure \ref{fisher_diag} this works remarkably well. It is computed 
entirely from the data, so one can argue that this is the actual 
curvature matrix needed in a Bayesian analysis, 
as opposed to its ensemble average, the Fisher matrix. 
This equation needs to be generalized for realistic surveys with a 
survey mask, gaps, variable noise etc. to its matrix form, 
but we will not pursue it here. 

\section{Appendix: Analysis in the presence of external additive nuisance parameters}
\label{appb}

Minimum variance power spectrum 
estimator in the form we presented in the main text has to be modified
when the data are contaminated or do not contain information for some
modes \cite{1992ApJ...398..169R}. For example, in LSS galaxy surveys the mean density is unknown and 
obtained from the measured data itself, which means that low $k$ modes
cannot be measured independently. The mean density mode needs to be assigned a large variance, which in turn forces its reconstructed value to zero. 
Because of the finite survey size a uniform 
density across the survey can be created from $\bi{k}>0$
modes: the window function of the $\bi{k}=0$ determines this mode mixing. 
Other contaminants exist as well, such as  
dust extinction, star-galaxy separation etc. Assuming 
we know their spatial structure, we can model them as a known 
data vector $\bi{L}(\bi{q})$, with an unknown amplitude $q$. 
One wants to make sure that the 
estimator remains unbiased and that the contaminated modes 
do not affect the power spectrum estimator. 

In some instances we want to remove these components from the data 
without knowing their actual values, in other cases we may be 
interested in their best reconstructed values. 
From the point of view of this paper we want to marginalize
over these modes  
by integrating over their probability distribution.
Let us discuss the case of simultaneous 
reconstruction and external parameter determination. Following 
\cite{1992ApJ...398..169R} and generalizing to the nonlinear case 
we can write the data vector
 in the presence of external parameters $\bi{q}$
as $\bi{d}=\bi{f(s)}+\bi{L}(\bi{q})+\bi{n}$, 
where $\bi{L}$ is the $N \times M_q$ nonlinear mapping
and $M_q$ is the number of external parameters. 
For example, when determining the mean density one can write $\bi{L}(q)=(1,1,...,1)q$ and if 
the data $\bi{d}$ express galaxy numbers on a pixelized grid $q$ corresponds to the mean number 
of galaxies per pixel. If the number density is evolving with redshift then one must determine the 
relative evolution from the model of the galaxy selection, 
so that only the overall normalization needs to be determined.  

We assume a gaussian
prior for $\bi{q}$ 
\begin{equation}
P_q(\bi{q}) = (2\pi)^{-M_q/2} \det(\bi{S_q})^{-1/2} \exp\left(-{1 \over 2}
\bi{q}^{\dag} \bi{S_q}^{-1}\bi{q}\right),
\end{equation}
with $\bi{S_q}$ the prior covariance matrix of the nuisance parameters. Often we have no prior knowledge 
of these nuisance parameters, in which case the prior is simply flat. 
The full posterior is 
\begin{equation}
P(\bi{q},\bi{s},\vert \bi{d}) = P_q(\bi{q})P_s(\bi{s})P_n[\bi{d}-
(\bi{f(s)}+\bi{L(q)})].
\end{equation}

We can now simultaneously maximize the posterior probability for 
$\bi{q}$ and $\bi{s}$. This gives 
$\hat{\bi{q}}$ 
and $\hat{\bi{s}}$, and the corresponding curvature matrix $\bi{D}$ now includes properly
the correlations between the nuisance parameters and the modes $\bi{s}$. 
If we want to know the probability distribution of the data independent
of external parameters and the underlying field we marginalize over
these parameters by integrating $P_q(\bi{q})P_s(\bi{s})P_n[\bi{d}-
(\bi{f(s)}+\bi{L(q)})]$ over $d^{M_q}\bi{q} d^M \bi{s}$,
as we did in equation \ref{marg}. The resulting posterior as a function of power spectrum $\bi{S}$ will properly 
account for all the correlations between the nuisance parameters and the power spectrum. 
Since our estimator is built out of a sequence of optimization steps, there is no 
difference to the equations we wrote above: we are simply minimizing $\chi^2$ 
for a few more parameters, with no significant cost increase given the large number of modes.  

When this is applied to 
the linear model one can write an explicit form of the covariance matrix, and with the use of Sherman-Morrison-Woodbury formula one can show that 
this corresponds to addition of $S_q\bi{LL}^{\dag}$ for each parameter $q$ to the covariance matrix of the data \cite{1992ApJ...398..169R}. 
However, one does not need to restrict to linear nuisance parameter models, nor 
does one need to write down the covariance matrix: since 
our procedure never requires an explicit form of the 
curvature matrix or its inverse, there is no need to know 
its form in the presence of these contaminated modes: optimization procedure and the corresponding construction of the
noise bias and Fisher matrix automatically give the desired 
result. 

\section{Appendix: 2-LPT Derivative}
\label{appc}

The optimization involves the evaluation of the derivative (equation 
\ref{gder}) term at each step
\be
\bi{g} = \bi{f'^{\dag}{N}^{-1}[d-f(s_m)}],
\ee
where we have ignored the derivative of the prior term for the time being and  $\bi{f'}$ is defined as the gradient of the forward model at every position with respect to every initial mode.

In the main text, we show the results using 2-LPT 
as the forward model. In this appendix we present the 
evaluation of its derivatives in this subsection. The corresponding 
derivatives at 1-LPT level have been presented in \cite{2013ApJ...772...63W} and at 2-LPT level in \cite{2013MNRAS.432..894J}, albeit in a different form from the one here. 
We use 2-LPT because it significantly improves the cross-correlation 
coefficient between the full nonlinear field and the approximate 2-LPT
field \cite{2012JCAP...12..011T}: in the limit of cross-correlation coefficient being unity the 
resulting solution multiplying 2-LPT with the transfer function becomes exact, and the 
error is given by the deviation of the cross-correlation coefficient from unity. As we discuss in the text, all of our 2-LPT fields are 
multiplied with the transfer function, which we will omit here for the 
sake of clarity. 
To make it more illustrative, we replace the generic notation used in the main text 
with more conventional notation for our specific model, i.e. we replace
the density modes of interest $\bi{s_m}$ with $\bi{\delta}_j(\bk)$ where the subscript $j$
is $0$ or $1$ for the real and complex parts of a Hermitian field respectively. We write the noise matrix as $N_{ij}=\sigma^2(\bi{r}_i)\delta^D_{ij}$. 
Thus the derivative explicitly summed over all the positions is 
\begin{eqnarray}
\label{fsum}
\bi{g}_j(\bk) & = & \sum_{\bx} \frac{\partial \bi{f_{\delta}}({\bx})}{\partial \bi{\delta}_j(\bk)}\frac{[\bi{d}(\bx) - \bi{f_{\delta}}({\bx})]}{\sigma^2(\bx)} \\
& = & \sum_{\bk_1} \frac{\partial \bi{f_\delta}^*(\bk_1)}{\partial \bi{\delta}(\bk)}\rho_d(\bk_1),
\end{eqnarray}
where $\rho_d(\bk_1) = \mathcal{F}\frac{[ \bi{d}(\bx) - \bi{f_{\delta}}({\bx})]}{\sigma^2(\bx)}$ 
and $\mathcal{F}$ is the Fourier transform ($\sum_x e^{-i\bk\cdot\bx}$). The forward model with CIC interpolation of particles from their Eulerian position $\br(\bq)$ is,
\be
\bi{f}^*_{\bi{\delta}}(\bk_1) = w(R_sk_1) \sum_\bx e^{i \bk_1\cdot \bx}\sum_{\bq} w_c(\bx - \br(\bq)) ,
\ee
 where $w(R_sk_1)$ is the smoothing kernel and $\sum_{\bq} w_c(\bx - \br(\bq))$ is the CIC kernel assignment. The 2 LPT forward model calculates the Eulerian positions of the particles to second order in Lagrangian displacement, $\br(\bq) = \bq + \bs^{(1)}(\bq) + \bs^{(2)}(\bq)$. These displacements in Fourier domain are 
\begin{eqnarray}
\bs^{(1)}(\bk) &=&  \frac{i\bk}{k^2}\delta(\bk), \\
\bs^{(2)}(\bk) &=& \frac{1}{2} \sum_{i \neq j}\sum_{\bk', \bk''} 
       \left( i\bk'_i i\bk''_j \bs^{(1)}_i(\bk') \bs^{(1)}_j(\bk'') - i\bk'_j i\bk''_i \bs^{(1)}_i(\bk') \bs^{(1)}_j(\bk'') \right) \times
       \nonumber \\ &\delta_D&(\bk - \bk' -\bk''). 
\end{eqnarray}
This second order displacement is simply the conventional definition \newline $\nabla \cdot \bs^{(2)}(\bq)=\frac{1}{2} \sum_{i \neq j} \left( \bs^{(1)}_{i, i} \bs^{(1)}_{j, j} - \bs^{(1)}_{i, j} \bs^{(1)}_{j, i} \right)$, written as convolutions in the Fourier space.

Thus, the complete derivative can be expressed as independent derivative of first and second order displacements
\begin{eqnarray}
g_j(\bk) 
& = & \sum_{\bk_1}\rho_d(\bk_1) w(R_sk_1)\sum_\bx e^{i\bk_1\cdot\bx} \sum_\bq
      \frac{ \partial w_c(\bx - \br(\bq))}{\partial \delta_j(\bk)} \\
& = & \sum_\bq \frac{\partial \br(\bq)}{\partial \delta_j(\bk)} \cdot \sum_{\bx} \frac{ \partial w_c(\bx - \br(\bq))}{\partial \br(\bq)} \sum_{\bk_1} \rho_d(\bk_1) w(R_sk_1) e^{i\bk_1\cdot\bx} \nonumber \\
& = & \sum_\bq \frac{\partial \bs^{(1)}(\bq)}{\partial \delta_j(\bk)} \cdot \Psi(\bq) + 
      \sum_\bq \frac{\partial \bs^{(2)}(\bq)}{\partial \delta_j(\bk)} \cdot \Psi(\bq)), \nonumber \\
& = & g_j^{(1)}(\bk) + g_j^{(2)}(\bk) ,
\end{eqnarray}
where, we have defined  
\begin{equation}
\Psi(\bq) = \sum_{\bx} \frac{ \partial w_c(\bx - \br(\bq))}{\partial \br(\bq)} \sum_{\bk_1} \rho_d(\bk_1) w(R_sk_1) e^{i\bk_1\cdot\bx},
\end{equation}
which can be evaluated using simple Fourier transforms and re-griddings.
Then, to first order (1-LPT), this leads to 
\begin{eqnarray}
g_j^{(1)}(\bk) 
& = & \sum_\bq \frac{\partial \bs^{(1)}(\bq)}{\partial \delta_j(\bk)} \cdot \Psi(\bq) \nonumber \\
& = & \sum_{\bk_2}  \frac{i\bk_2}{k_2^2} \frac{\partial \delta(\bk_2)}{\partial \delta_j(\bk)}\cdot \sum_\bq e^{i\bk_2\cdot\bq}\Psi(\bq) \nonumber \\
& = & \sum_{\bk_2}  \frac{i\bk_2}{k_2^2} \frac{\partial \delta(\bk_2)}{\partial \delta_j(\bk)}\cdot \Psi^*(\bk_2)
\end{eqnarray}
which leads to the final expression for the most modes
\begin{eqnarray}
g_0^{(1)}(\bk) & = & (2) \frac{\bk}{k^2}\cdot \Psi_1(\bk)\\
g_1^{(1)}(\bk) & = & - (2) \frac{\bk}{k^2}\cdot \Psi_0(\bk).
\end{eqnarray}
while  the factor of $(2)$ drops out for the self-conjugating (zero and the nyquist) modes of the hermitian field. 

At second order (2-LPT), we have
\begin{eqnarray}
g_j^{(2)}(\bk) 
& = & \sum_\bq \frac{\partial \bs^{(2)}(\bq)}{\partial \delta_j(\bk)} \cdot \Psi(\bq) \\
& = & \sum_{\bk_2} \sum_{i \neq j}\sum_{\bk', \bk''} - \frac{i\bk_2}{k_2^2} \frac{1}{2}  
      \frac{\left( \bk_i^{\prime  2} \bk_j ^{\prime \prime 2}  - \bk'_j \bk''_i \bk'_i \bk''_j  \right)}{ k^{\prime  2} k^{\prime \prime 2}} 
      \frac{\partial ( \delta(\bk') \delta(\bk'') \delta_D(\bk_2 - \bk' -\bk'')) }{\partial \delta_j(\bk)} \cdot \Psi^*(\bk_2) \nonumber \\
& = & - \sum_{\bk_2} \sum_{i \neq j}\sum_{\bk', \bk''}\frac{\left( \bk_i^{\prime 2} \bk_j^{\prime \prime 2}  - \bk'_j \bk''_i \bk'_i \bk''_j  \right)}{ k^{\prime  2} k^{\prime \prime 2}} 
      \delta(\bk') \delta_D(\bk_2 - \bk' -\bk'') \frac{\partial \delta(\bk'')}{\partial \delta_j(\bk)}
      \frac{i\bk_2}{k_2^2} \cdot \Psi^*(\bk_2), \nonumber
\end{eqnarray}
where in the first expression we use result from first order derivative and in the second step we are using the symmetry between $\bk' $ and $\bk'' $. As done above for $j= 0$, we have:
\begin{eqnarray}
g_0^{(2)}(\bk) 
& = & - \sum_{\bk_2} \sum_{i \neq j}\sum_{\bk', \bk''}\frac{\left( \bk_i^{\prime 2} \bk_j ^{\prime \prime 2}  - \bk'_j \bk''_i \bk'_i \bk''_j  \right)}{k^{\prime  2} k^{\prime \prime 2}} 
      \delta(\bk') \delta_D(\bk_2 - \bk' -\bk'')     \times \nonumber \\ 
      && \left( \delta_D(\bk'' - \bk) +  \delta_D(\bk'' + \bk) \right)
      \frac{i\bk_2}{k_2^2} \cdot \Psi^*(\bk_2) 	\nonumber  \\
& = & 2Re \left[ \sum_{i \neq j} \left( \frac{\bk_j ^2}{k^2}\sum_{\bk_2} \frac{ (\bk_2 -\bk)_i^2 }{(k_2 - k)^2} \delta (\bk - \bk_2)
      \frac{i \bk_2}{k_2^2} \cdot \Psi(\bk_2) \right. \right. \nonumber \\
     && \left. \left. -  \frac{\bk_i\bk_j }{k^2}\sum_{\bk_2} 
      \frac{ (\bk_2 -\bk)_i (\bk_2 -\bk)_j}{(k_2 - k)^2} \delta (\bk - \bk_2)\frac{i \bk_2}{k_2^2} \cdot \Psi(\bk_2 \right) \right],
\end{eqnarray}
where simplifications have been made using symmetry between different dummy variables 
and Hermitian properties of $\delta$ and $\Psi$.

The final expression is a convolution that can be evaluated as a product in the configuration space after defining 6 $\alpha$ functions and a $\beta$ function as:
\begin{eqnarray}
\alpha_{ij}(\bx) & = & \mathcal{F}^{-1}[\frac{\bk'_i\bk'_j }{k'^2} \delta(\bk')], \\
\beta(\bx) & = & \mathcal{F}^{-1}[\frac{i \bk' }{k'^2} \Psi(\bk')],
\end{eqnarray}
leading to the second order terms 
\begin{eqnarray}
g_0^{(2)}(\bk) 
& = & 2Re \left[ \sum_{i \neq j} \left( \frac{\bk_j ^2}{k^2} \mathcal{F}[\alpha_{ii}(\bx)\beta(\bx)] - 
      \frac{\bk_i\bk_j }{k^2} \mathcal{F}[\alpha_{ij}(\bx)\beta(\bx)] \right) \right], \\
g_1^{(2)}(\bk) 
& = & 2 Im \left[ \sum_{i \neq j} \left( \frac{\bk_j ^2}{k^2} \mathcal{F}[\alpha_{ii}(\bx)\beta(\bx)] - 
      \frac{\bk_i\bk_j }{k^2} \mathcal{F}[\alpha_{ij}(\bx)\beta(\bx)] \right) \right].
\end{eqnarray}

\bibliographystyle{revtex}
\bibliography{cosmo,cosmopp}

\end{document}